\documentclass[13pt]{article}
  \usepackage[english]{babel}
\usepackage{amsthm}
\usepackage{amsmath}

\numberwithin{figure}{section}
\usepackage{amsfonts}
\usepackage{graphics}
\usepackage{graphicx}
\usepackage{adjustbox}
\usepackage{color}
\usepackage{makecell}
\usepackage{pstricks}
\usepackage{ulem}
\usepackage[latin1]{inputenc}
\usepackage{soul}
\usepackage{url}
\usepackage{xcolor} 
\usepackage{graphics}
\usepackage{graphicx} 
\usepackage{setspace}
\usepackage{footnote}
\usepackage{natbib}

\newfont{\spp}{msbm10 scaled \magstep1}
\newfont{\sppp}{msbm7 scaled \magstep1}

\newtheorem{remark}{Remark}

\newtheorem{definition}{Definition}

\date{\vspace{-5ex}}
\begin{document}

\title{\vspace{-2cm}\bf \Large Can market volumes reveal market makers' rationality and a new risk premium?}%

\author{
Francesca Mariani \\ \small{Dipartimento di Scienze Economiche e Sociali, Universit\`a Politecnica delle Marche, Italy}\\
\small{Ph. n. +39 071 2207243, Fax n. +39 071 2207102, E-mail: f.mariani@staff.univpm.it} \\
Maria Cristina Recchioni\\ 
\small{Dipartimento di Scienze Economiche e Sociali, Universit\`a Politecnica delle Marche, Italy}\\ 
\small{Ph. n. +39 071 2207066 , Fax n.+39 071 2207102, E-mail: m.c.recchioni@staff.univpm.it} \\
Tai-Ho Wang \\
\small{Baruch College, The City University of New York, New York}\\
\small{Ph. +01 646 3123997, Fax n. +01 646 3124111  E-mail: tai-ho.wang93@login.cuny.edu}\\
Roberto Giacalone\\
\small{Global Head of Liquid Strategies, Icona Capital - London, UK,  E-mail: r$_-$giacalone@rocketmail.com }
}

\maketitle

\begin{abstract}
Based on empirical findings, we introduce a fully observable model for price-volume dynamics to estimate market friction and trading price impact. The model determines the impacted price dynamics as the dynamics ensuring the observed dollar volume to be the solution to an utility problem in a frictionless market for a representative  riskless principal market maker {which operates simultaneously in both risky and risk-free assets.}  Interestingly, the model enables to measure the market risk aversion and the market price for the trading risk, which results around 2.5 as prescribed by Prospect Theory. The impacted price dynamics demonstrates strong forecasting potential, as shown in the online appendix. Extensive empirical analyses support the model. 
\end{abstract}

\noindent \textbf{Data Availability}\\
The data that support the findings of this study are available in Yahoo Finance at https://finance.yahoo.com/. These data were derived from the following resources available in the public domain: \\
- Market indices, https://finance.yahoo.com/quote/\\
- Exchange rates:, https://finance.yahoo.com/markets/currencies/

\noindent \textbf{Conflict of Interest}\\
The authors have no conflicts of interest to declare.

\section{Introduction}\label{intro}

The starting point of this contribution is an extensive empirical analysis conducted on price-volume dynamics of different risky assets and the idea of interpreting  {these} dynamics as the optimal strategy of a representative riskless principal market maker. \textcolor{red}{Hereafter, the term  trading volume refers to the instantaneous flow of shares traded per unit of time, representing the stochastic rate of market activity rather than its cumulative sum.} Our empirical findings suggest lognormal models for both price and volume with time-dependent parameters (see Eqs. (\ref{dSt})-(\ref{def_corr})) able to explain market friction and trading price impact.
Specifically, estimating the model parameters from real data, we found that the trading volume drift-to-diffusion ratio of very liquid and well-established risky assets remains constant over time and equal to approximately 2.5, while, this ratio deviates from this value when estimated from emerging assets. We interpret this deviation as a measure of market friction. To capture the trading price impact, we micro-found our model by interpreting the observed dollar trading volume  as the  aggregated ``optimal" activities of riskless principal market makers, who operate in a frictionless market.
This aggregated ``optimal" dollar trading volume is 
the optimal strategy of a representative riskless principal market maker, who operates in a frictionless market for a suitable shadow price (see Eq. (\ref{def_ipf})). The dynamics of the shadow price includes the parameters of the trading volume dynamics, while revealing the impact of the trading activity on price.

Specifically, we consider non-proprietary (or principal-less or riskless principal) market makers who intermediate order flows across two markets: a risky asset and a risk-free asset 
with the goal of maximizing the expected commissions. 
Unlike traditional (or principal) market makers, the riskless principal market makers do not use their own capital to hold assets in their balance sheet overnight as at the end of any trading day their net position is always zero. Their activity is limited to facilitate transactions between buyers and  sellers, without ever carrying any market exposure to the following day. Their profit does not come from the price changes of the asset, but only from the commissions they earn, which are fees proportional to the total dollar volume intermediated. That is,  the expected revenue of each riskless principal market maker is a percentage of the total dollar volume\footnote{The total dollar volume is the sum between the risky dollar volume and the risk-free dollar volume. The risky dollar volume is the total value of shares or contracts traded, calculated as (number of shares traded) x (price per share). The risk-free dollar volume is defined similarly.}.

To model the activity of riskless principal market makers, we  formulate  a dynamic control problem, where the state variable is the total dollar volume and the control variable is the risky dollar volume,
while the utility function is  a HARA function.
We refer to this problem as Optimal Riskless Principal Market Making (ORPMM) problem at micro level.
Under suitable assumptions on price and volume dynamics parameters for each market maker, we prove that the aggregated risky dollar volume  dynamics is the solution of an  ORPMM problem at macro level for a suitable price (see Eq. (\ref{def_ipf})). Hereafter, for simplicity, we refer to the ORPMM problem at macro level simply as ORPMM problem.  In case of trading activity and market frictions, our approach transfers the frictions to the asset price, revealing their impact on price drift and volatility. 
This impact is due to two sources of market frictions. 
The first is the imperfect correlation between the two Brownian processes of price and volume (also known as decoupling), which affects both drift and volatility. 
The second originates from the  drift of volume and it is related to the variance-adjusted risk premium\footnote{The variance-adjusted risk premium is the difference between the risk premium and the variance of the asset price, i.e., $\mu-r-\sigma^2.$}, but it affects only the price drift. 
Thus, trading an asset in a market with frictions impacts not only price drift, as commonly discussed in the literature, but also price volatility.

\subsection{Main results}

We begin by assuming the following price-volume dynamics for the risky asset:
\begin{align}
\frac{d S_t}{S_t}=&\mu dt +\sigma dZ_t,\quad t\in [0,T],\label{dSt}\\
\frac{dn_t}{n_t}=&\eta(\psi\,dt +\sigma\,dQ_t),\quad t\in[0,T],\label{def_vol}
\end{align} 
where $\mu$ and $\sigma$ are the drift and volatility of the asset price, $\psi$ is the volume drift, while $Z_t$, $Q_t$ are Brownian motions and $dZ_t$, $dQ_t$ their differentials and $\rho$ is the correlation coefficient:
\begin{equation}\label{def_corr}
E(dZ_t,dQ_t)=\rho dt.
\end{equation}
Note that, a positive value of $\rho$ implies that positive price shocks tend to be associated with positive volume shocks (see \cite{Clark1973}).
In (\ref{def_vol}) $\eta$ is a scaling factor that controls the overall speed and volatility of the volume dynamics. We refer to $\eta$ as \textit{trading intensity}.
A high value of $\eta$ 
leads to faster growth and greater fluctuations in trading volume, while a low value of $\eta$ indicates a more stable volume trajectory. 

Model (\ref{dSt})-(\ref{def_corr})  is in line with the findings  of  \cite{BCCM2018} and  \cite{BCCMR2019}, which  suggest that market makers face two risks: one due to asset performance driven by the asset Brownian motion, and a second due to information circulating in the market, which can modify market makers' beliefs. 
Moreover, model (\ref{dSt})-(\ref{def_corr}) aligns with  the interpretation provided in \cite{Clark1973} and in \cite{Dallores2017}, according to which both price movements and trading volumes are driven by a common, unobservable latent process.
However,  while strongly correlated, volume and price dynamics are not the same process (see \cite{Tauchen1983}). The empirical analyses presented in Section \ref{sec5} also support the validity of our model. 
 {Evidence of approximately lognormal behavior in daily stock market trading volume was pioneered by \cite{AJ1989}. This empirical regularity aligns with the theoretical foundations of the Mixture-of-Distributions Hypothesis (\cite{Clark1973}) and finds further support in later behavioral models, such as those of \cite{O1998} and \cite{O1999}. In particular, investor overconfidence, by inducing heterogeneous and multiplicative trading responses to private information, provides a plausible behavioral micro-foundation for the observed lognormality of volume, a thesis central in  \cite{BO2001} and \cite{STV2006}. }

We propose a micro-founded model that focuses on a representative riskless principal market maker, whose activity  represents the aggregated riskless principal market-making activity. This representative riskless principal market maker directly affects the dynamics of traded volume.
Specifically, in Eq. (\ref{def_vol}) the trading intensity $\eta$  is an endogenous variable directly managed by the representative riskless principal market maker. It represents the intensity of incoming buy and sell orders that the representative riskless principal market maker manages. We assume that the representative riskless principal market maker solves the ORPMM problem.

However, while the dynamics (\ref{dSt})-(\ref{def_corr})  accurately describes the observed price and volume dynamics, its parameters, when calibrated with real data, are inconsistent with the micro-founded model. 
Specifically, according to this model, given $r$ the risk-free rate, the volume drift $\psi$ should be equal to $\mu-r-\sigma^2$ and, consequently, the drift-to-diffusion ratio of  volume process $\psi/\sigma$ should be equal to $(\mu-r-\sigma^2)/\sigma$ (as illustrated in Section \ref{sec3}), but this value is not consistent with real data.

We interpret the observed misalignment  between the observed drift-to-diffusion ratio of volume and the theoretical one as the signal of presence of inefficiencies in the market.

The idea is to transfer the friction observed in the volume dynamics into the price through the ORPMM problem.
Specifically, we expect that, in case of no trading ($\eta = 0$), the trading price is the  {efficient} price.
Otherwise, in case of an increasing trading activity ($\eta > 0$), the representative riskless principal market maker has to handle a growing number of orders. This requires them to actively adjust bid and ask prices to manage the order flow. The  {riskless principal} market maker's actions inevitably move the market price, causing a \textit{price impact} that grows with the trading intensity, $\eta$. 
Market friction directly amplifies price impact. This implies that for a given trading intensity ($\eta$), a greater degree of friction will lead to a larger price impact. More specifically, the market's level of friction acts as a leverage multiplier on the price impact generated by trading.
For example, in a market with little friction, a high value of trading intensity $\eta$ is needed to cause even a small price movement. Vice versa, in a market with a lot of friction, even a small value of trading intensity $\eta$ can produce a significant price impact. 

We prove the existence of a price $\tilde S_t$ ( {the impacted price}) such that the observed dynamics of the risky dollar volume, i.e., $\pi_t=n_tS_t$, corresponds to the optimal dynamics of the ORPMM problem in a frictionless market for the asset price $\tilde S_t$ (see Section \ref{sec4}).

The dynamics of the impacted price is proven to be:
\begin{equation}\label{def_ipf}
\frac{d \tilde S_t}{\tilde S_t}=\left(\mu+\frac{(f-\sigma^2(1-\rho))\eta}{1+\eta}\right)dt+\tilde\sigma dW_t,\quad t\in [0,T],
\end{equation}
with 
\begin{equation}\label{sig_vol}
\tilde\sigma=\sigma\frac{\sqrt{1+\eta^2+2\rho\eta}}{1+\eta},
 \end{equation}
where in (\ref{def_ipf}) $W_t$ is a Brownian motion obtained from $Z_t$ and $Q_t$  (see Section \ref{sec4}), $dW_t$ its differential,  while $\rho$ is the correlation coefficient in (\ref{def_corr}).
In Eqs. (\ref{def_ipf})-(\ref{sig_vol}),  $f$ is the deviation due to the friction of the  drift volume from optimal one in a frictionless market, that is:
 \begin{eqnarray}\label{ft}
  {f=\psi-(\mu-r-\sigma^2)}.
\end{eqnarray}

In Eq. (\ref{def_ipf}) the term  $\frac{(f-\sigma^2(1-\rho))\eta}{1+\eta}$ is the impact on the price drift, it consists of two terms: one  the misalignment of the volume drift from the theoretical one ($f\neq 0$), the second one the misalignment of price and volume Brownian motions ($\rho\neq 1$). If we choose  $f=0$ (i.e., we remove the misalignment of the volume drifts) and $\rho=1$ (i.e., we remove the misalignment of the Brownian motions) the price drift reduces to $\mu.$

The term $\frac{\sqrt{1+\eta^2+2\rho\eta}}{1+\eta}$ is the impact on the price volatility and it depends only from the misalignment of price and volume Brownian motions ($\rho\neq 1$). 

If we choose  $\rho=1$ (i.e., we remove the misalignment of the Brownian motions) we get $\tilde\sigma=\sigma.$

Eq. (\ref{def_ipf})  defines a new market impact model, according to which, in presence of friction, the trading affects the price generating a new source of noise ($W$) and a change in the price drift and volatility.
 
As shown in Section \ref{sec5}, the parameters in the dynamics (\ref{def_ipf})-(\ref{sig_vol}) are fully observable: $\mu,$ $\sigma$ and $\rho$ are estimated first, followed by $\psi$ from the ratio drift-to-diffusion of the volume, then $f$  using Eq. (\ref{ft}) and, finally, $\eta$.
The dynamics of the impacted price reveals how the trading intensity  ($\eta$) affects price drift and volatility in presence of frictions ($f\neq 0$ and $\rho\neq1$). 

Interestingly,  the trading intensity  $\eta$  plays the role of a perturbative parameter. When $\eta$ approaches zero (i.e., negligible volume fluctuations) the price dynamics $d\tilde S_t/\tilde S_t$ reduces to the dynamics of the no impacted price (\ref{dSt}). Conversely, as $\eta$ approaches infinity the impacted volatility, $\tilde\sigma$, converges to the non-impacted one, $\sigma$, while the price drift converges to $r+(\psi+\rho\sigma^2)$. This indicates that, in this extreme case, the impacted price growth depends solely on the parameters of the volume dynamics.

As mentioned above, the price-volume equations (\ref{def_ipf})-(\ref{sig_vol}) are proven to be the unique dynamics under which the observed total dollar volume corresponds to the  optimal strategy of a representative riskless principal market maker that solves the ORPMM problem in a frictionless market for the asset price $\tilde S_t$, $t>0$. As a byproduct,  we can recover the market risk aversion and the market price of risk associated with  this dynamics.

The risk aversion parameter is given by (see Section \ref{sec4} for further details):
\begin{eqnarray}\label{ugamma_def}
1-\bar{\gamma}=\frac{(\mu-r)(1+\eta)+(f-\sigma^2(1-\rho))\eta}{\sigma^2(1+\eta^2+2\rho\eta)}.
\end{eqnarray}

The market price of risk of the impacted price, that is will refers ``trading price of risk", is given by  (see  Section \ref{sec4}):
\begin{equation}\label{def_mkt}
Trading\, Price\, of\, risk=\frac{\xi^p+\eta\,\xi^v}{\sqrt{1+\eta^2+2\rho\eta}}=\frac{(\mu-r)(1+\eta)+(f-\sigma^2(1-\rho))\eta}{\sigma\sqrt{1+\eta^2+2\rho\eta}},
\end{equation}
where $\xi^p$ is the market price of risk and $\xi^v$ is the volume price of risk. Therefore, as expected, the impacted price combines two sources of risk: the market price of risk $\left(\xi^p=\frac{\mu-r}{\sigma}\right)$ and the volume price of risk $\left(\xi^v=\frac{\mu-r}{\sigma}+\frac{f-\sigma^2(1-\rho)}{\sigma}\right)$ (see Section \ref{sec4} for more details). The combined market price of risk reduces to the well-known market price of risk $\xi^p$ when the friction $f$ is zero and the correlation coefficient is one ($\rho = 1$). That is, when there is no trading friction and the two Brownian motions are perfectly positively correlated.

Concluding, we summarize the  threefold contribution of the paper.

Firstly, we examine whether the dynamics of asset volume exhibits features consistent with those outlined by the theoretical ORPMM model developed in Section \ref{sec3}. To accomplish this, we assume that the volume is a diffusion process and we calculate the drift-to-diffusion ratio of volume process for  {different} asset classes. Our estimations indicate that this ratio remains relatively stable over extended time periods, with minor fluctuations observed across various asset classes. However, notably, the estimated  {fractions} diverge significantly from  {zero}, except for market indices such as the S\&P500 and Nasdaq. This finding suggests the presence of market inefficiencies.

Secondly, assuming that the volume serves as a proxy for the total dollar volume intermediated by a representative riskless principal market maker in the risky asset  {and using the price-volume dynamics (\ref{dSt})-(\ref{def_corr})}, we derive the dynamics of the total dollar volume invested in the aforementioned asset. The drift and diffusion of the total dollar volume dynamics are direct functions of both the asset price's drift and diffusion, as well as the trading intensity.

The crucial assumption is that the risky dollar volume  strategy $\pi_t$ {(see Eq. (\ref{dpit_def}))} is optimal for the financial market, as it represents the strategy derived from the ORPMM problem for an ``ideal asset price" that accounts for market frictions in both its drift and volatility. {This assumption is supported by the micro-founded model of Section \ref{sec2}.} The ideal price corresponds to what we defined as the ``impacted price" in the Introduction (see Eq. (\ref{def_ipf})). The key idea is to transfer the frictions observed in the total dollar volume dynamics to the price, making that dynamics optimal in the ORPMM frictionless market for a representative riskless principal market maker with trading intensity $\eta$.
\\
As a byproduct, we can express market risk aversion as a function of the observable parameters in price and volume dynamics and can estimate it.

Finally, the approach used to derive the impacted price suggests that trading produces two sources of market friction. The first source originates from the misalignment of volume drift with that prescribed by ORPMM dynamics, i.e., the friction $f$ in Eq. (\ref{def_ipf}). The second originates from the misalignment of the Brownian processes of price and volume. Thus, trading influences price dynamics not only through the price drift, as commonly discussed in the literature on market impact, but also by affecting price volatility, which is due to the second source of friction.

The two sources of friction generate a new price of risk that can be estimated using both price and volume data. Surprisingly, it closely aligns with what is outlined by Prospect Theory, as illustrated in \cite{KT1979} and  \cite{KT1991} (see Section \ref{sec3}). We refer to this as the trading price of risk because it measures the extra return that market makers demand to bear risk in the presence of trading. Naturally, the trading price of risk reduces to the market price of risk  { when $\rho=1$ (only one source of noise) and $f=0$ (no frictions)}. Interestingly, in this case, the drift-to-diffusion ratio of volume dynamics coincides with that of the ORPMM model.

\subsection{Related scientific literature}

Our work is related to different strands of scientific literature.

Firstly, the literature that deals with the analysis and modeling of trading volumes as a dynamic process. 
In the scientific literature many measures for trading volume are proposed (see \cite {Karpoff1987} and \cite{Lo2009} for a review). 
The main measures are the total number of shares traded, the total value of the shares traded (dollar volume), the number of individual transactions, or the normalized measured given by the turnover, and the wealth turnover.
In \cite{Lo2009} the authors use the turnover as  measure of trading volume. The turnover is defined as the number of shares traded divided by the total number of shares outstanding.
Starting from some empirical stylized facts, the authors identify several key properties of trading volume, including its high autocorrelation and persistence, and its strong positive correlation with the asset price volatility. {They propose a dynamic general equilibrium model where the sole driver of volume dynamics is the activity of the investors rebalancing their portfolio. The trading volume is assumed to follow a mean-reverting Ornstein-Uhlenbeck diffusion process that successfully reproduces the stylized facts.}
The main merit of the paper is to propose a new vision of volume for which 
the volume is not a random byproduct of trading but a process with its own predictable dynamics that can reveal  market parameters, such as investor risk aversion, the market depth and the presence of market frictions.
{\cite{Guasoni2017} use wealth turnover as primary measure of trading volume.} The wealth turnover is defined as the total dollar amount traded per unit of time, divided by the investor's total wealth. The paper proposes a model to derive trading volume dynamics in a market context with frictions. Unlike models that treat price and volume as separate processes, this work establishes a direct analytical link. The dynamics of volume (OU) and price (geometric Brownian motion) are connected through the investor's action. The investor continuously rebalances their portfolio in response to price shocks, and this rebalancing generates the trading volume. Price impact, which makes instantaneous rebalancing costly, forces the investor to split orders, giving rise to a stochastic and persistent volume dynamic.
{In \cite{Clark1973} the volume is measured as number of total number of traded shares and  the Mixture of Distributions Hypothesis (MDH) theory is introduced.} This theory hypothesizes that both the asset price and the volume are driven by an unobservable process, such as the flow of new information into the market. 
More recently, \cite{Dallores2017} extend the classic MDH theory proposing a dynamic model in which the  price and volume are explained by two unobservable (latent) variables: the arrival rate of information and the level of liquidity frictions, which captures market imperfections such as bid-ask spread or order execution costs. Interestingly, in this model the level of liquidity frictions acts as a modulator of trading volume. When frictions are high, trading becomes more costly. 
In \cite{Wang2024} an extensive empirical analysis on the relationship between price and volume in the Bitcoin Futures ETF market is conducted. The analysis reveals a strong positive correlation between price volatility and trading volume  and suggests that volume can be used as a valid indicator of information flow in the market. 
{The role of volumes in determining the profitability of a market is analyzed in \cite{Menkvelda2013}. In this paper the market maker is assumed to operate  in both a primary (incumbent) market and an emerging (entrant) market.   He controls the trading speed to provide liquidity taking advantage of differences in fees and order flow. }
{An other way to measure of volume is the dollar volume. Dollar trading volume refers to the total value of an asset traded over a given period. It is computed by multiplying the number of shares traded by the price of each share. Differently from the share volume, that limits to count  the total number of exchanged shares, the dollar volume reflects the financial value of trades and is particularly useful for comparing the trading activity of assets with different prices.}
{In \cite{Caginalp2017} the authors analyze the relationship between dollar volume and market inefficiency. According to their study, as dollar volume increases, the price efficiency (measured as the deviation between the ETF's trading price and its net asset value) increases. However, as volume overcomes a certain threshold, the relationship reverses. Price efficiency decreases as dollar volume gets even higher.}
{Moreover, the dollar volume is a useful tool to measure trading cost. In \cite{Goyenko2024} the authors use dollar volume as a direct proxy for market liquidity and postulates  an inverse relationship between price and dollar volume. According to the principles outlined in \cite{Goyenko2024},  minimizing trading costs requires maximizing the dollar volume.} 
{In \cite{Chung2004} the dollar trading volume is used as explicative variable in a regression to explain the price spread. The authors find that the dollar volume impacts negatively on the price spread.} 

Secondly, the literature on the impact of trading on asset prices. Models for price impacts have been widely studied; for a survey of them, refer to  \cite{B2017}. \cite{HM2005} address the problem of liquidating a large block of risky asset shares in the presence of price impact and transaction costs.  \cite{C2019} empirically examines whether trade-size clustering, which measures trading irregularity, impacts price formation.  \cite{G2010} investigates the relationship between the shape of the market impact function and the decay of market impact.
\textcolor{blue}{This strand of literature has recently evolved to incorporate the perspective of liquidity providers. 
\cite{CW2020}  investigate how the impact of trading on price dynamics can be anticipated using \textit{alpha signals}, allowing market makers to adjust their quotes proactively.
\cite{J2021} explores optimal market making under persistent order flow, specifically accounting for the long-term price impact of trades and the resulting inventory risk.
\cite{BHS2025} further extend this by introducing a model with exogenous competition, analyzing how the aggregate inventory of competitors influences price stability and execution probabilities.
Our work builds upon these foundations by expanding the research to the study of riskless principal market maker behavior, a specific strategic configuration that remains less explored in the current literature.}

Thirdly, the literature that explores the role of liquidity and frictions in financial markets using microstructure models for the behavior of financial investors. In \cite{KPZ2024} the authors examine the role of liquidity in competition among ETFs modeling the interactions between high-turnover and low-turnover investors. The model explains why liquidity attracts high-turnover investors, justifying higher fee. In \cite{ABR2024} the authors integrate market frictions (such as transaction costs and lending constraints) into a dynamic equilibrium framework to understand their impact on asset pricing, trading activity, and market efficiency. \cite{HKN2024} use a network-based model to explore how brokerage networks facilitate liquidity and manage the impact of large trades in financial markets. In \cite{FMVV2022} a standard noisy rational expectations model is used to investigate the implications of data availability and information flow in financial markets, focusing on how investors optimally gather and use information in the face of uncertainty and limited data. In \cite{GIJR2024} the authors use a dynamic stochastic general equilibrium (DSGE) model to capture the relationship between credit cycles, expectations, and corporate investment.

Lastly, the literature on Prospect Theory by  \cite{KT1991}. The central assumption of this theory is that losses and disadvantages have a greater impact on preferences than gains and advantages. The implications of loss aversion for economic behavior are considered. Kahneman and Tversky propose empirical experiments suggesting that a loss aversion coefficient of about 2-2.5 may explain both risky and riskless choices involving monetary outcomes and consumption goods. Surprisingly, the price of risk experienced from trading volume shows the same value for assets with a long life in the financial market.

\subsection{Paper outline}
The paper is structured as follows: Section \ref{sec2} presents the empirical finding and, starting from the observed volume dynamics, introduces a model for the risky dollar volume. Section \ref{sec3} presents the Optimal Riskless Principal Market Making (ORPMM) problem and solves it. Section \ref{sec4}   proposes an interpretation for the model introduced in Section \ref{sec2} based on  a representative riskless principal market maker describing the aggregate behavior of rational heterogeneous riskless principal market makers who solve their ORPMM problem. {As a byproduct, the dynamics of the impacted price is derived, as well as the market risk aversion and the trading price of risk.} Section \ref{sec5} proposes an empirical analysis where the experienced risk premium and the market risk aversion are estimated.

\section{Empirical finding and implications}\label{sec2}
We start from empirical findings obtained by looking the time series of volumes and prices of different assets. For simplicity, we assume that the asset price process is a geometric Brownian motion as in Eq. (\ref{dSt}).
We estimate the volatility of the price process, $\sigma$ using a year of consecutive price daily.

Further, we  empirically observe:
\begin{enumerate}
\item[$a)$] The volume process, $n_t$, $t>0$, is a diffusion  given by:
\begin{equation}\label{dnt}
dn_t=n_t(\mu^vdt +\sigma^vdQ_t),\quad t\in[0,T],
\end{equation} 
where $Q_t$, $t>0$, is a Brownian motion while  $dQ_t$ is its differential;

\item[$b)$] The ratio, $\mu^v/\sigma^v$ (i.e., drift-to-diffusion) of volume process  is a piecewise constant function of time.

\item[$c)$] The ratio $\mu^v/\sigma^v$ multiplied by $\sigma$ (i.e., the price volatility) is a piecewise constant function with respect to time, so  implying the existence of a relationship between  volume volatility and asset volatility. {Therefore, without loss of generality, assuming the volume drift as a linear function of trading intensity,} we can rewrite the drift and the diffusion terms as follows:
$\mu^v=\eta\psi$, $\sigma^v=\eta\sigma$. The dynamics of the volume looks like:
\begin{equation}\label{dvol_def}
dn_t=n_t\eta(\psi\,dt +\sigma\,dQ_t),\quad t\in[0,T].
\end{equation} 

\item[$d)$] The Brownian motions which define  the asset-price and volume processes are correlated but with a weakly correlation. That is, we have
\begin{equation}\label{corr}
E(dZ_t,dQ_t)=\rho dt,\quad t\in[0,T],
\end{equation}
where $\rho$ is close to zero.

\end{enumerate}
Note that, except for the assumption that volume drift depends linearly on the trading intensity, the volume model (\ref{dvol_def}) is completely data-driven.
The main consequence of dynamics (\ref{dSt}), (\ref{dnt}) and (\ref{corr}) is the dynamics for the risky dollar volume, $\pi_t=n_tS_t,$ that reads:
\begin{equation}\label{dpit}
d\pi_t=\pi_t\left[\left(\mu + (\psi+\rho\sigma^2)\eta\right)dt +\sigma(dZ_t+\eta dQ_t)\right],\quad t\in[0,T].
\end{equation}
Note that the  risky dollar volume, i.e., $\pi_t$, given in Eq. (\ref{dpit}), depends
on two noises, $Z_t,$ related to the fundamentals of the asset $S_t$, while $Q_t$ is related to the market makers' trading and beliefs.

In this market, the no-arbitrage principle holds, thus, there exists a change of  measure that allows the discounted processes $e^{-rt}S_t$, $e^{-rt}\pi_t$ to be martingales. Here $r$ is the risk free interest rate.
This change of measure, which ensures the absence of arbitrage in pricing, is given by:
\begin{eqnarray}\label{cm}\label{dZ}
dZ_t&=&d\widetilde{Z}_t-\frac{(\mu-r)}{\sigma}dt,\quad t\in[0,T],\\
dQ_t&=&d\widetilde{Q}_t-\frac{(\psi+\rho\sigma^2)\eta}{\eta\sigma}dt\, \quad t\in[0,T].\label{dQ}
\end{eqnarray}
The change of measure reveals the presence of two risks. The market price of risk $\xi^p=\frac{(\mu-r)}{\sigma}$  measures the extra gain in terms of  price volatility units that a market maker requires to take on the price risk compared to investing in a guaranteed income security.  The volume price of  risk:
\begin{equation}\label{risk_v}
 \xi^v=\frac{\eta(\psi+\sigma^2\rho)}{\eta\sigma}=\frac{(\psi+\sigma^2\rho)}{\sigma}
\end{equation} 
measures the extra gain in terms of volume volatility units that a market maker requires to take on the risky dollar volume $\pi_t$  compared to investing in a strategy consisting only of guaranteed income security, which is well-know and interpretable as  the compensation for each unit of volatility risk.

We can rewrite $\pi_t$ using a new Brownian motion which is the linear combination of the previous two as follows:
\begin{equation}\label{dpit_def}
\frac{d\pi_t}{\pi_t}=\left(\mu + (\psi+\rho\sigma^2)\eta\right)dt +\sigma\sqrt{1+\eta^2+2\rho\eta}dW_t,\quad t\in[0,T].
\end{equation}

\section{Optimal Riskless Principal Market Making  problem at micro level}\label{sec3}

We consider a  non-proprietary (or riskless principal) market maker who intermediates in the time interval $[0,T]$ (the investment period) order flows across two markets: a risky asset with price $S_t$ and a risk-free asset with price $B_t$.   

The risk-free asset $B_t$ earns a constant rate of return $r$, while the risky asset price $S_t$ follows a geometric Brownian motion with diffusion term is driven by the Brownian motion which drives the risky dollar volume $\pi_t$ (see, Eq. (\ref{dpit_def})), that is:
\begin{equation}\label{rev1}
    \frac{dS_t}{S_t}= \mu dt + \sigma dW_t,\quad t\in[0,T].
\end{equation}
The drift and volatility in Eq. (\ref{rev1}) are those in Eq. (\ref{dSt}).

A  riskless principal market maker operates with the sole purpose of maximizing their total profit. Unlike traditional market makers, he does not use his own capital to hold assets, thus eliminating inventory risk. His activity is constrained by the principle of cash-neutrality: every transaction they facilitate is a ``pass-through" between a buyer and a seller, without ever taking on a net position.
	
The market maker's profit does not come from asset price changes, but only from the commissions they earn, which are directly proportional to the total dollar volume of the transactions he intermediates.

Let $n_t$ and $n_t^0$ denote the corresponding traded quantities.  The total dollar volume (i.e., the total notional processed by the market maker) is given by
\begin{equation}
	x_{t} = n_t \, S_t + n_t^B \, B_t,\quad t\in[0,T].
\end{equation}
The market maker's revenue is assumed to be proportional to $x_{t}$, reflecting a commission-based or spread-capture remuneration scheme as observed in  Electronic Communication Networks (ECNs), crypto exchanges, or agency market makers.

We impose the following condition on dollar volume dynamics:
\begin{equation}
		dx_{t} =  n_t \, dS_t + n^B_t\, dB_t , \quad t\in[0,T],
\end{equation}
which is equivalent to
\begin{equation}
	S_t \, dn_t + B_t \, dn^B_t +d[n, S]_t = 0, \quad t\in[0,T].
\end{equation}
	
We call this the \textit{Notional Conservation Constraint}.  Economically, this assumption captures the idea that the {total flow of orders} available to the market maker is exogenously determined and cannot be increased at will.  The market maker, being riskless principal, does not provide additional balance-sheet capacity to create or absorb trading volume. Instead, he can only influence how the existing order flow is {reallocated} between the risky and the risk-free market.  
	
This constraint ensures that variations in total dollar volume arise solely from price fluctuations in the underlying assets, rather than from exogenous changes in traded quantities. It reflects the notion of {flow preservation}:  The intermediary cannot generate new notional, but can alter its {composition} across venues by adjusting incentives, spreads, or routing policies.  
	
In practice, a riskless principal market maker earns fees proportional to notional intermediated.  His revenues do not depend on inventory mark-to-market fluctuations, but on the total activity he manages. 

 {Consequently, the economic levers available do not involve direct control over $n_t$ or $n_t^0$, but rather consist of indirect mechanisms designed to shift the distribution of trading activity. These levers include the adjustment of fee structures and rebates to favor specific markets, the management of spreads and order book depth to attract order flow, and the strategic routing of orders across various venues to concentrate activity in desired locations.}
	
The Notional Conservation Constraint captures this reality by abstracting from endogenous flow creation  and focusing on the {allocation problem}: given a fixed pool of trading activity, 
how should it be distributed across risky vs.\ risk-free venues?  

The Notional Conservation Constraint and the price dynamics (\ref{rev1}) imply that the dynamics of the total dollar volume  solves:
\begin{equation}\label{nnq}
    dx_t = \left[rx_t + \pi_t(\mu - r) \right]dt + \pi_t \sigma \, dW_t, \quad t\in[0,T],
\end{equation}
where $\pi_t=n_t S_t$ denotes the risky dollar volume, i.e. the dollar volume allocated to the risky asset.

The market maker's goal is to maximize  the expected utility of  {terminal total dollar volume}  $x_T$ by controlling the risky dollar volume $\pi_t=n_t S_t$, subject to the conservation constraint. Formally, the Optimal Riskless Principal Market Making (ORPMM) problem can be formulated  as
\begin{equation}
	\max_{\pi_t} \; \mathbb{E}\big[ U(x_{T}) \big],
\end{equation}
where $U(\cdot)$ is  a hyperbolic absolute risk aversion (HARA) utility function of the form
\begin{equation}
   U(x) = \frac{1-\gamma}{\gamma}\left(\frac{ax}{1-\gamma}\right)^\gamma, \quad a,\gamma>0, 
\end{equation}
Here, $\gamma$ represents the coefficient of relative risk aversion  of the market maker. 
 
Economically, this means the market maker aims to steer the flow composition towards the risky market 
whenever price--volume correlations make this allocation more profitable. Since revenues scale with notional, concentrating activity in states where price movements are favorable  increases expected commission income.

The optimal risky dollar volume, solution of ORPMM problem, is given by
\begin{equation}\label{nnw}
    \pi_t^* = \frac{\mu - r}{\sigma^2(1-\gamma)}x_t, \quad t\in[0,T].
\end{equation}

The fraction between the optimal risky dollar volume and the total dollar volume is constant over time.  It reflects the fundamental trade-off between risk and return: the excess return $\mu-r$ scaled by the market maker's relative risk aversion $\gamma$ and the variance $\sigma^2$ of the risky asset.

The optimal risky dollar volume behavior is the same observed in the dynamics of the trader inventory  in a frictionless Merton framework (see \cite{Merton1971}, \cite{MCR2019} for details on the share dynamics and \cite{Herdegen2021} for the verification theorem.

Using Eq. (\ref{nnq}) and applying the \^Ito lemma to $\pi_t^*$ in Eq. (\ref{nnw}), we have:
\begin{equation}\label{nnr}
\frac{d\pi_t^*}{\pi_t^*}=\left(r+\frac{(\mu-r)^2}{\sigma^2(1-\gamma)}\right) dt+\frac{\mu-r}{\sigma(1-\gamma)}dW_t, \quad t\in[0,T].
\end{equation}
On the other hand, the \^Ito lemma  and the definition of $\pi_t^*$ ($\pi_t^*=n_t^* S_t$) imply:
\begin{equation}\label{nnr1}
d\pi_t^*=n^*_t dS_t+ S_t dn_t^*+ d[n^*,S]_t
\end{equation}
where $d[\cdot,\cdot]$ denotes the bracket process (see \cite{Protter2004}). Assuming $n^*_t$ and $S_t$ positively perfectly correlated, bearing in mind the dynamics of $S_t$ (Eq. (\ref{rev1})), equating Eqs. (\ref{nnr}), (\ref{nnr1}) we have:
\begin{equation}
\frac{dn_t^*}{n^*_t}= \frac{\mu-r}{\sigma^2(1-\gamma)}((\mu-r-\sigma^2)dt+\sigma dW_t),\quad t\in[0,T]. 
\end{equation}
\section{Optimal consumption and market risk aversion}\label{sec4}

In this section, we interpret the empirical findings from Section \ref{sec2} within the framework of a representative riskless principal market maker to address two delicate issues:
a) is the market risk aversion parameter measurable from observed data? b) can the volumes reveal the rationality of market makers?

Hereafter, we will refer to the riskless principal market maker simply as the market maker.

To this end, we propose a collective model for the dynamics of risky dollar volume. In particular, we demonstrate that this dynamics results from the collective dynamics of heterogeneous market makers with different risk aversions whose beliefs about the price vary but who make decisions on their dollar volume according to their ORPMM optimal solution. 
 
Specifically, in our model,  each market maker acts rationally solving his ORPMM problem. As a consequence, the optimal risky dollar volume, derived as a solution of the  ORPMM problem, associated with different market makers are different.  Assuming  market makers are independent and share the same distribution of the risk aversion and the same distribution of the risk-adjusted price of risk, we show that, when the number of market makers is large enough, the individual  dollar volume dynamics of market makers are  independent copies of the same process under different  risk aversions and risk-adjusted price of risk. This allows us to derive  the {collective dynamics of the risky dollar volume} as the limit dynamics of the individual {dynamics} when the number of market makers goes to infinity. 
{The collective dynamics}  is given by a   diffusion process with drift and volatility expressed in terms of the market risk aversion and of the mean and variance of the risk-adjusted price of risk.

Finally, {the collective dynamics of risky dollar volume}  is interpreted as the {strategy} of an ``ideal"  market maker with  risk aversion given by the market risk aversion, who operates in a frictionless  market on an ``ideal"  risky asset, whose price drift and volatility depend on the mean and variance of the risk-adjusted price of risk.  This ideal market maker is the fictitious representative market maker that represents the multitude of market makers at the macroscopic level  when the number of market makers is large enough. {The ideal risky asset is the impacted price announced in the Introduction.}

{In this section, for the sake of simplicity, we use the word ``strategy to denote the dynamics of risky dollar volume resulting from the solution of the  ORPMM problem.}

In what follows, we propose a mathematical formalization of our model.

We have $N$ market makers, each of them has a different belief about the risky asset price drift and volatility and a different risk aversion.

Denoting by $S_{i,t},$ $t\in[0,T],$ the price associated with the $i-$th market maker, for $i=1,2,\ldots,N,$ we assume that $S_{i,t}$ are \^Ito processes, evolving according to the stochastic differential equation system
\begin{eqnarray}\label{ipdn}
\frac{d S_{i,t}}{S_{i,t}}=\mu_{i} dt +\sigma_{i}dW_{i,t},\quad t\in [0,T],
\end{eqnarray}
where  $W_{i,t},$   $t\in[0,T],$ $i=1,2,\ldots,N,$   are  $N$ independent  standard Brownian motions. 

In Eq. (\ref{ipdn}) the quantities $\mu_{i}$ and $\sigma_{i},$   are, respectively, the return and volatility of the price  perceived by the $i-$th market maker, $i=1,2,\ldots,N.$

We denote by $B_t$ the price of the risk-free asset at time $t,$ $t\in [0, T],$ which satisfies the following deterministic differential equation:
\begin{eqnarray}\label{irf}
dB_t = rB_t dt, t \in [0, T], 
\end{eqnarray}
where $r > 0$ is a risk-free interest rate, and $B(0) = 1.$ In other words, all market makers  share the same view on the risk free return.

Given $n^B_{i,t}$ and $n_{i,t}$ the number of shares invested, respectively, in the risk-free and risky assets, and the price $S_{i,t}$ at time $t,$ $t\in[0,T],$ the total dollar volume of the market maker $i,$ $x_{i,t},$ $t\in[0,T],$ reads:
\begin{eqnarray}\label{ixt}
x_{i,t}=n_{i,t}S_{i,t}+n^B_{i,t} B_t,\quad t\in [0,T], \, i=1,2,\ldots,N.
\end{eqnarray}
Moreover, assuming the Notional Conservation Constraint, the dynamics of $x_{i,t}$ is given by
\begin{eqnarray}\label{ixtdn}
d x_{i,t}=[x_{i,t} r+n_{i,t}S_{i,t}(\mu_{i}-r)]dt+n_{i,t}S_{i,t}\sigma_{i} dW_{i,t},\quad t\in [0,T], \, i=1,2,\ldots,N,
\end{eqnarray}
Now, denoting by $\pi_{i,t}=n_{i,t}S_{i,t},$ $t\in[0,T],$  Eq. (\ref{ixtdn}) can be rewritten as
\begin{eqnarray}\label{i6}
d x_{i,t}-rx_{i,t}dt=\pi_{i,t}\left[(\mu_{i}-r)dt+\sigma_{i} dW_{i,t}\right],\quad t\in [0,T],\, i=1,2,\ldots,N,
\end{eqnarray}
with the initial condition $x_{i,0}=\hat x_{i,0}.$

For $i=1,2,\ldots,N,$ the optimal solution of the ORPMM problem with HARA utility function for the i-th market maker (see Section \ref{sec3})  is:
\begin{eqnarray}\label{i10}
\pi_{i,t}=(1+\eta_{i})\, x_{i,t},  \quad t\in[0,T],\, i=1,2,\ldots,N,
\end{eqnarray}
where
\begin{eqnarray}\label{vel}
\eta_{i}=\frac{\mu_{i}-r}{\sigma_{i}^2(1-\gamma_{i})}-1, \quad  i=1,2,\ldots,N,
\end{eqnarray}
is the trading intensity of the $i-$th market maker  and $\gamma_{i}\in\mathbb R$ is the risk aversion parameter  of the $i-$th market maker at time $t$ 

The dynamics (\ref{i10}) is the  dynamics of rational heterogeneous market makers {that implement ORPMM strategies}.

Using Eq. (\ref{i6}) and applying the \^Ito lemma to $\pi_{i,t},$ in (\ref{i10}), we have:
\begin{eqnarray}\label{ipi_opt}
\displaystyle \frac{d\pi_{i,t}}{\pi_{i,t}}=&\displaystyle \left(r +\alpha_{i}^2(1-\gamma_{i})\right) dt+\alpha_{i}dW_{i,t},\quad t\in[0,T],\, i=1,2,\ldots,N,
\end{eqnarray}
or, equivalently,
\begin{eqnarray}\label{a4}
d\ln \pi_{i,t}=\left(r+\displaystyle\alpha_{i}^2\left(\frac{1}{2}-\gamma_{i}\right)\right)\, dt+\alpha_{i}dW_{i,t}, \quad t\in[0,T],\, i=1,2,\ldots,N,
\end{eqnarray}
where: 
\begin{eqnarray}\label{alpha}
\alpha_{i}=\frac{\mu_{i}-r}{\sigma_{i}(1-\gamma_{i})},\quad i=1,2,\ldots,N.
\end{eqnarray} 
 
Note that the quantity in (\ref{alpha}) is the risk-adjusted price of risk, i.e., the price of the risk that takes into account the market maker's individual preference for the risk. A higher value of the risk aversion parameter indicates greater risk aversion, reducing the impact of the excess return on the overall risk measure. In other words, a market maker with higher risk aversion will require a higher excess return to compensate for the additional risk, thereby reducing the price of risk value.

For $i=1,2,\ldots,N,$ each market maker $i$ has its individual  risk aversion, $\gamma_{i}.$ This assumption, together with the assumption that each market maker has a different belief on asset return and volatility, implies that each market maker has different values of the risk-adjusted price of the risk (\ref{alpha}). 

\begin{remark}\label{central}{\textbf{Main Assumption.}} \textit{The market makers have different risk aversions and  beliefs on price drift and volatility, i.e., the price drift and volatility are random variables such that  the corresponding risk-adjusted prices of risk are drawn from the same distribution and the risk aversions are random variables drawn from the same distribution. More specifically, we assume  that, for  $i=1,2,\ldots,N,$  the quantities $\alpha_{i}$ 
are Independent and Identically Distributed (IID) variables with mean $\mu_{\alpha}$ and variance $\sigma^2_{\alpha}.$\\
Analogously, the risk aversion parameters $\gamma_{i}$ are IID variables with mean $\bar{\gamma}$, and variance $\sigma_{\gamma}.$ 
Moreover, we assume that the random variables $\alpha_{i}$ and $\gamma_{i}$ are independent and {mutually} independent of the Brownian motions $W_{i,t},$ $t\in[0,T],$ $i=1,2,\ldots,N.$
}
\end{remark}

We determine the aggregate dynamics for the market makers when their number $N$ goes to infinity. To this end, once fixed $t\in [0,T]$, we first discretize the market makers' dynamics as usual:
\begin{align}\label{a4d}
&&\ln \pi_{i,t+\Delta t}-\ln\pi_{i,t}=\left(r+\displaystyle\alpha_{i}^2\left(\frac{1}{2}-\gamma_{i}\right)\right)\, \Delta t+\alpha_{i}{ {\Delta W_{i}}}, \quad i=1,2,\ldots,N,
\end{align}
{where:
\begin{equation}\label{deltaW}
  {\Delta W_{i}}=W_{i,t+\Delta t}-W_{i,t}, \quad i=1,2,\ldots,N.
\end{equation}}
For later convenience, we set:
\begin{equation}\label{xii}
 {\xi_{i,\Delta t}}=\ln \pi_{i,t+\Delta t}-\ln \pi_{i,t}, \quad i=1,2,\ldots,N.
\end{equation}

\begin{remark}\textit{We show that  for any fixed $t$, $t\in (0,T)$ and for small enough $\Delta t$, the sample  $ {(\xi_{i,\Delta t})_{i=1,2,\ldots,N}},$   is drawn from the same population $ {\xi_{\Delta t}}$ and we define the  {collective} strategy  $\tilde{\pi}_{t}$ as follows:}
\end{remark}

\begin{definition}\label{def1}
\textit{Let $ {\xi_{\Delta t}}$ be the population that {draws} the market makers' dynamics. We define the  {collective} strategy $\tilde{\pi}_t$, associated with $ {\xi_{\Delta t}}$, as the {solution of}  the following discrete-time first-order difference equation:}
\begin{equation}\label{tildep}
 \ln(\tilde{\pi}_{t+\Delta t}) - \ln(\tilde{\pi}_t) =  {\xi_{\Delta t}}.
\end{equation}
\end{definition}
{Now, following  Dai et al. (2023) (see Appendix A in \cite{DDJZ2023}),  in Eq. (\ref{a4d})  we replace $\alpha_{i}$ with $\mu_{\alpha}+\sigma_{\alpha}\varepsilon_{i},$ where  {$\varepsilon_{i}$ is a random variable with zero mean and unit variance independent of $W_{i,t},$} and we replace $\gamma_{i}$ with $\bar{\gamma}+\sigma_{\gamma}\varphi_{i},$ where  {$\varphi_{i}$ is a random variable with zero mean and unit variance independent of $W_{i,t}.$} Note that, the independence assumption of $\alpha_{i}$ and $\gamma_{i}$ implies that $\varepsilon_{i}$ and $\varphi_{i}$ are independent. Then, for $i=1,2,\ldots,N$, we can rewrite Eq. (\ref{a4d}) as follows:
 \begin{align}\label{a4e}
 {\xi_{i,\Delta t}}=&\left(r+\displaystyle(\mu_{\alpha}+\sigma_{\alpha}\varepsilon_{i})^2\left(\frac{1}{2}-(\bar{\gamma}+\sigma_{\gamma}\varphi_{i})\right)\right)\, \Delta t+(\mu_{\alpha}+\sigma_{\alpha}\varepsilon_{i})\Delta W_{i,t} \nonumber\\&=\left(r+(\mu_{\alpha}^2+\sigma_{\alpha}^2)\left(\frac{1}{2}-\bar{\gamma}\right)\right)\Delta t+(\mu_{\alpha}+\sigma_{\alpha}\varepsilon_{i})\Delta W_{i,t}+res_{i},\end{align}
where the residual term, $res_{i},$ is given by:
\begin{align}\label{res}
res_{i}=&2\mu_{\alpha}\sigma_{\alpha}\varepsilon_{i}\left(\frac{1}{2}-\bar{\gamma}\right)\Delta t-\sigma_{\gamma}\varphi_{i}(\mu_{\alpha}^2+\sigma_{\alpha}^2+2\mu_{\alpha}\sigma_{\alpha}\varepsilon_{i})\Delta t\nonumber\\&+\sigma_{\alpha}^2(\varepsilon_{i}^2-1)\left(\frac{1}{2}-\bar{\gamma}\right)\Delta t.
\end{align}
Using the Law of Large Numbers, recalling that $\varepsilon_{i}$ and $\varphi_{i}$ are mutually independent random variables with zero mean and unit variance for $i=1,2,\ldots,N$, and bearing in mind the independence between  $\varepsilon_{i}$ and $\varphi_{i}$, $i=1,2,\ldots,N$, it is simple to verify that:
\begin{eqnarray}\label{lim_b}
\frac{1}{N}\sum_{i=1}^N res_{i}\xrightarrow[N\to +\infty]{\text{a.s.}} 0 \quad \text{ and  }\quad \frac{1}{N}\sum_{i=1}^N res^2_{i}\xrightarrow[N\to +\infty]{\text{a.s.}} o(\Delta t),\quad \Delta t\to 0^+.
\end{eqnarray}
In Eq.(\ref{lim_b}) the symbol a.s. stands for ``almost surely'' and $o(\cdot)$ is the Landau symbol.

Note that the random variables $ {\xi_{i,\Delta t}},$  $i = 1,2,\ldots,N,$ are transformations of the IID normal random variables $\varepsilon_{i},$ $\varphi_{i}$ and $ {\Delta W_{i}},$ then  $ {\xi_{i,\Delta t}},$  $i = 1,2,\ldots,N,$ are IID normal random variables sampled from an unknown population $ {\xi_{\Delta t}}$.
Using the Law of Large Numbers, we recover the mean and variance of the population $  {\xi_{\Delta t}}$, respectively,  as the limit for $N\to +\infty$  of the sample mean and of the sample variance of $ {\xi_{i,\Delta t}},$ $i=1,2,\ldots,N$ (computed as the difference between the  second order sample moment and the square of the sample mean).

The sample mean of $ {\xi_{i,\Delta t}}$ is given by:
\begin{align}\label{sample_mean}
\frac{1}{N}\sum_{i=1}^N  {\xi_{i,\Delta t}}=& \left(r+(\mu_{\alpha}+\sigma_{\alpha})^2\left(\frac{1}{2}-\bar{\gamma}\right)\right)\Delta t\nonumber\\&+\mu_{\alpha}\frac{1}{N}\sum_{i=1}^N  {\Delta W_{i}}+\sigma_{\alpha}\frac{1}{N}\sum_{i=1}^N\varepsilon_{i} {\Delta W_{i}}+\frac{1}{N}\sum_{i=1}^N res_{i}.
\end{align}

Analogously, the second order sample moment of  $\Delta \ln\pi_{i,t}$ is:
\begin{eqnarray}\label{sec_mom}
\frac{1}{N}\sum_{i=1}^N ( {\xi_{i,\Delta t}})^2
= (\mu_{\alpha}+\sigma_{\alpha})^2\Delta t+o(\Delta t), \quad\quad \Delta t\to 0^+.
\end{eqnarray} 

Using the Law of Large Numbers and  Eqs. (\ref{lim_b}) and  taking the limit as $N\to +\infty$ of (\ref{sample_mean}) and (\ref{sec_mom}) we can compute the mean and variance of the population $ {\xi_{\Delta t}}$ as follows:
\begin{eqnarray}\label{mean_pop}
\frac{1}{N}\sum_{i=1}^N  {\xi_{i,\Delta t}} \xrightarrow[N\to +\infty]{\text{a.s.}}\left(r+(\mu_{\alpha}+\sigma_{\alpha})^2\left(\frac{1}{2}-\bar{\gamma}\right)\right)\Delta t=\mathbb E[ {\xi_{\Delta t}}],
\end{eqnarray}
\begin{eqnarray}\label{var_pop}
\frac{1}{N}\sum_{i=1}^N ( {\xi_{i,\Delta t}})^2-\left(\frac{1}{N}\sum_{i=1}^N  {\xi_{i,\Delta t}}\right)^2 \xrightarrow[N\to +\infty]{\text{a.s.}} (\mu_{\alpha}+\sigma_{\alpha})^2\Delta t+o(\Delta t)=Var[ {\xi_{\Delta t}}],\quad \Delta t\to 0^+.
\end{eqnarray}
Now, observing that $\varepsilon_{i} {\Delta W_{i}}$ are  {mutually} independent random variables with zero mean and unit variance, the Central Limit Theorem implies that  $\frac{1}{N}\sum_{i=1}^N \varepsilon_{i} {\Delta W_{i}}$ converges, as $N\to +\infty,$ almost surely to a standard normal random variable $W_t$. 
Therefore, using Eqs. (\ref{mean_pop}), (\ref{var_pop}) we have the following:
\begin{eqnarray}\label{a5e}
  {\xi_{\Delta t}}=\left[r+(\mu_{\alpha}^2+\sigma^2_{\alpha})\left(\frac{1}{2}-\bar{\gamma}\right)\right]\, \Delta t+\sqrt{\mu_{\alpha}^2+\sigma_{\alpha}^2}  {Z_{\Delta t}},
\end{eqnarray}
where $ {Z_{\Delta t}}$ is a normal random variable with zero mean and variance $\Delta t$ for any fixed $t$ and  $\Delta t$.  We rewrite $ {Z_{\Delta t}}$ as the difference of a Brownian process  evaluated at $t$ and $t+\Delta t$, that is, $ {Z_{\Delta t}}=W_{t+\Delta t}-W_t$, and we use the definition of the {``collective strategy"} in   Eq. (\ref{tildep}) for $ {\xi_{\Delta t}}$:  

\begin{eqnarray}\label{a5e}
 \ln(\tilde{\pi}_{t+\Delta t})- \ln(\tilde{\pi}_{t})=\left[r+(\mu_{\alpha}^2+\sigma^2_{\alpha})\left(\frac{1}{2}-\bar{\gamma}\right)\right]\, \Delta t+\sqrt{\mu_{\alpha}^2+\sigma_{\alpha}^2}(W_{t+\Delta t}-W_t),\quad t\in[0,T].
\end{eqnarray}
Finally, taking the limit for $\Delta t\to 0^+$ we obtain:
\begin{eqnarray}\label{a5ee}
d \ln \tilde \pi_{t}=\left[r+(\mu_{\alpha}^2+\sigma^2_{\alpha})\left(\frac{1}{2}-\bar{\gamma}\right)\right]\, d t+\sqrt{\mu_{\alpha}^2+\sigma_{\alpha}^2}dW_t,\quad t\in[0,T].
\end{eqnarray}

From Eq. (\ref{a5ee}), using the \^Ito lemma, we obtain the dynamics of $\tilde \pi_t$:
\begin{eqnarray}\label{a5d}
\frac{d \tilde\pi_{t}}{\tilde\pi_t}=\left[r+(\mu_{\alpha}^2+\sigma^2_{\alpha})\left(1-\bar{\gamma}\right)\right]\, dt+\sqrt{\mu_{\alpha}^2+\sigma_{\alpha}^2}dW_t,\quad t\in[0,T].
\end{eqnarray}
Eq. (\ref{a5d}) is the {collective dynamics of market makers acting in the market}.

Comparing Eqs. (\ref{ipi_opt}), (\ref{a5d}) we note that, as the number $N$ of market makers goes to infinity, the  process $\tilde \pi_t$ has a dynamics of the same form of $\pi_{i,t}$ where $\alpha_{i}$ is substituted by $\sqrt{\mu_{\alpha}^2+\sigma_{\alpha}^2}$ and $\gamma_{i}$ by $\bar{\gamma}$. 

This suggests that the collective dynamics (\ref{a5d}) can be interpreted as the dynamics of a representative market maker with risk aversion $\bar\gamma$ who implements an ORPMM optimal strategy with risk-adjusted price of risk $\sqrt{\mu_{\alpha}^2+\sigma_{\alpha}^2}.$ As a consequence, there exists an ideal asset price $\tilde S$ such that:
\begin{eqnarray}
d\tilde S_t=\tilde S_t (\tilde\mu dt+\tilde \sigma dW_t), \quad t\in[0,T],
\end{eqnarray} 
and 
\begin{eqnarray}\label{nn}
\frac{\tilde\mu-r}{\tilde\sigma(1-\bar\gamma)}=\sqrt{\mu_{\alpha}^2+\sigma_{\alpha}^2},\quad t\in[0,T].
\end{eqnarray}

Substituting Eq. (\ref{nn}) into dynamics (\ref{a5d}) we obtain:
\begin{eqnarray}\label{a5dd}
\frac{d \tilde\pi_{t}}{\tilde\pi_t}=\left[r+\frac{(\tilde\mu-r)^2}{\tilde\sigma^2(1-\bar\gamma)}\right]\, dt+\frac{\tilde\mu-r}{\tilde\sigma(1-\bar\gamma)}dW_t,\quad t\in[0,T],
\end{eqnarray}
or, equivalently, by rewriting Eq. (\ref{a5dd}) in terms of ORPMM, that is, setting $1+\tilde\eta=(\tilde \mu - r)/(\tilde\sigma^2(1-\overline{\gamma})^2)$:
\begin{eqnarray}\label{a5ddd}
\frac{d \tilde\pi_{t}}{\tilde\pi_t}=r\, dt+(1+\tilde\eta)\left[(\tilde\mu-r)\, dt+\tilde\sigma\, dW_t\right],\quad t\in[0,T],
\end{eqnarray}

where $\tilde\eta$ is the ideal trading intensity  of the representative market maker. 

Finally, imposing that the  ``ideal" dynamics of risky dollar volume (\ref{a5ddd}) mimics the observed dynamics  (\ref{dpit_def}) and that the ideal trading intensity $\tilde\eta$ is equal to the 
observed trading intensity $\eta$, we obtain:
\begin{eqnarray}\label{eq}
\left\{\begin{array}{l}
\displaystyle \tilde\eta=\eta,\\
\displaystyle (1+\tilde\eta)(\tilde\mu-r)= \mu-r+(\psi+\rho\sigma^2)\eta,\\
\displaystyle (1+\tilde\eta)\tilde\sigma=\sigma\sqrt{1+\eta^2+2\rho\eta}.
\end{array}\right.
\end{eqnarray}
%
From Eqs. (\ref{eq}) we can recover the drift and volatility of the impacted price, as follows:
\begin{align}\label{prop_e31}
\tilde\mu=&r+\frac{\mu-r+(\psi+\rho\sigma^2)\eta}{1+\eta},\\
\tilde\sigma =&\frac{\sigma}{1+\eta}\sqrt{1+\eta^2+2\rho\eta}.\label{prop_e32}
\end{align}

Furthermore,  from Eqs. (\ref{eq}) we can recover the market risk aversion  $\bar{\gamma}$  as follows:
\begin{eqnarray}\label{gamma}
\bar{\gamma}=1-\frac{\mu-r+(\psi+\rho\sigma^2)\eta}{\sigma^2(1+\eta^2+2\rho\eta)}.
\end{eqnarray}
Eq. (\ref{gamma}) allows us to estimate the market risk aversion.

Finally, using Eqs. (\ref{prop_e31}), (\ref{prop_e32}) we derive the ``trading price" of the risk announced in the Introduction (see formula (\ref{def_mkt})), i.e., the Sharpe ratio of the impacted price :
\begin{eqnarray}\label{priceRisk}
Trading \, Price\,Risk = \frac{\tilde \mu-r}{\tilde \sigma}=\frac{\mu-r+(\psi+\rho\sigma^2)\eta}{\sigma \sqrt{1+\eta^2+2\rho\eta}}.
\end{eqnarray}

The trading price of risk measures the extra gain in terms of units of the impacted price volatility that
market makers require to take on the strategy risk compared to investing in a strategy consisting only
of guaranteed income security. Note that the trading price of risk makes the observed discounted total dollar volume, defined in (\ref{dpit_def}), a martingale.

We emphasize that the trading price of risk prescribed by our model reduces to the classical market price of risk, $(\mu - r)/\sigma$, when the friction $f$ is zero and $\rho$ is equal to one. That is, as prescribed by ORPMM model, when there is no trading friction and the two Brownian motions are perfectly positively correlated.

\section{Empirical Analysis}\label{sec5}

We now illustrate the aforementioned findings. We consider different asset classes. Specifically, we examine the daily prices and the corresponding volumes of two stocks, Pfizer (PFZ) and Verizon Communications (VZ), from February 28th, 2018, to April 28th, 2022. Additionally, we analyze two market indices, S\&P 500 and NASDAQ, from September 17th, 2018, to September 15th, 2023. 
We use the Treasury Bill with a maturity of 3 months as the risk-free rate over the same period as the assets considered.

In the following of this section we model all the assets by using the dynamics (\ref{dSt})  and  the dynamics (\ref{dnt}) for the corresponding volumes.
 
We estimate the daily time series of the parameters $\mu$ and $\sigma$ of the risky asset  using a calibration window of about six months (i.e., 125 consecutive trading days)  and, in the last experiment, a window of 4 years (i.e., one year of 252 consecutive trading days). We move the window along the time series discarding the oldest observation and inserting the newest one. The subscript $t$ indicates the last date of the time-window used to estimate not only $\mu$ and $\sigma$ but also  the parameters $\mu^v,$ $\sigma^v$ of the Geometric Brownian Motion (GBM) describing the dynamics of the trading volume and the drift-to-volatility ratio, $\psi$, of volume as well as the correlation between the two Brownian processes $\rho$. These time series of the estimated parameters allow us to compute  the trading intensity $\eta$ along with the market risk aversion  $\bar{\gamma}$ (i.e., Eq. (\ref{gamma})) and the ``observed price of risk". The aforementioned parameters are considered constant throughout the time window.

The main finding is that for all asset classes the estimated value of $\bar{\gamma}$ is a constant function of time and its value is approximately 0.5. This is a very surprising result that supports those about the price of risk associated with the trading volume.

{Specifically, the observed ratio $\psi/\sigma=\mu^v/\sigma^v$ is a piecewise constant function of time. 
Interestingly, for stocks that are well-established in the financial market, this ratio is estimated approximately to be around 2.7. This value is very close to 2.5 that is the value prescribed by the Prospect Theory (see \cite{KT1991} page 1054).}

We now present the above-mentioned findings focusing on different asset classes.

 For each asset class, we begin by examining the extent to which the dynamics outlined by Eqs. (\ref{dSt}) and (\ref{dnt}) is true for both prices and volumes.
 We expect the log-normality of trading volumes
 while the log-normality of the corresponding prices is expected to fail.

Subsequently, among other estimations, we calculate the ratio of drift-to-diffusion for volumes, the volume and market price of risk, the trading intensity, and the risk aversion parameter.

We conclude with a section comparing these parameter values across different asset classes.

\subsection{PFE and VZ Stocks}\label{PFE_VZ}
Figure \ref{KS_Test_PFE_VZ} displays the p-values (on the $y$-axis) obtained from the Kolmogorov-Smirnov test, where the null hypothesis assumes normality of the time series. This test is applied to both the log-volume (upper panels) and the log-price (lower panels) for Pfizer and Verizon Communications within each window utilized in the estimation process. The $y$-axis ranges from 0.05 to 1.

We observe that the p-values for the log-volume are consistently greater than or equal to 0.05 for all dates, indicating that the null hypothesis cannot be rejected. However, in the case of log-prices, we find that the null hypothesis is rejected on several occasions. Specifically, for PFE and VZ stock prices, normality is particularly rejected during the Covid years (2020-2021).
\begin{center}
\begin{figure}[h!]
\centering\includegraphics[scale=0.35]{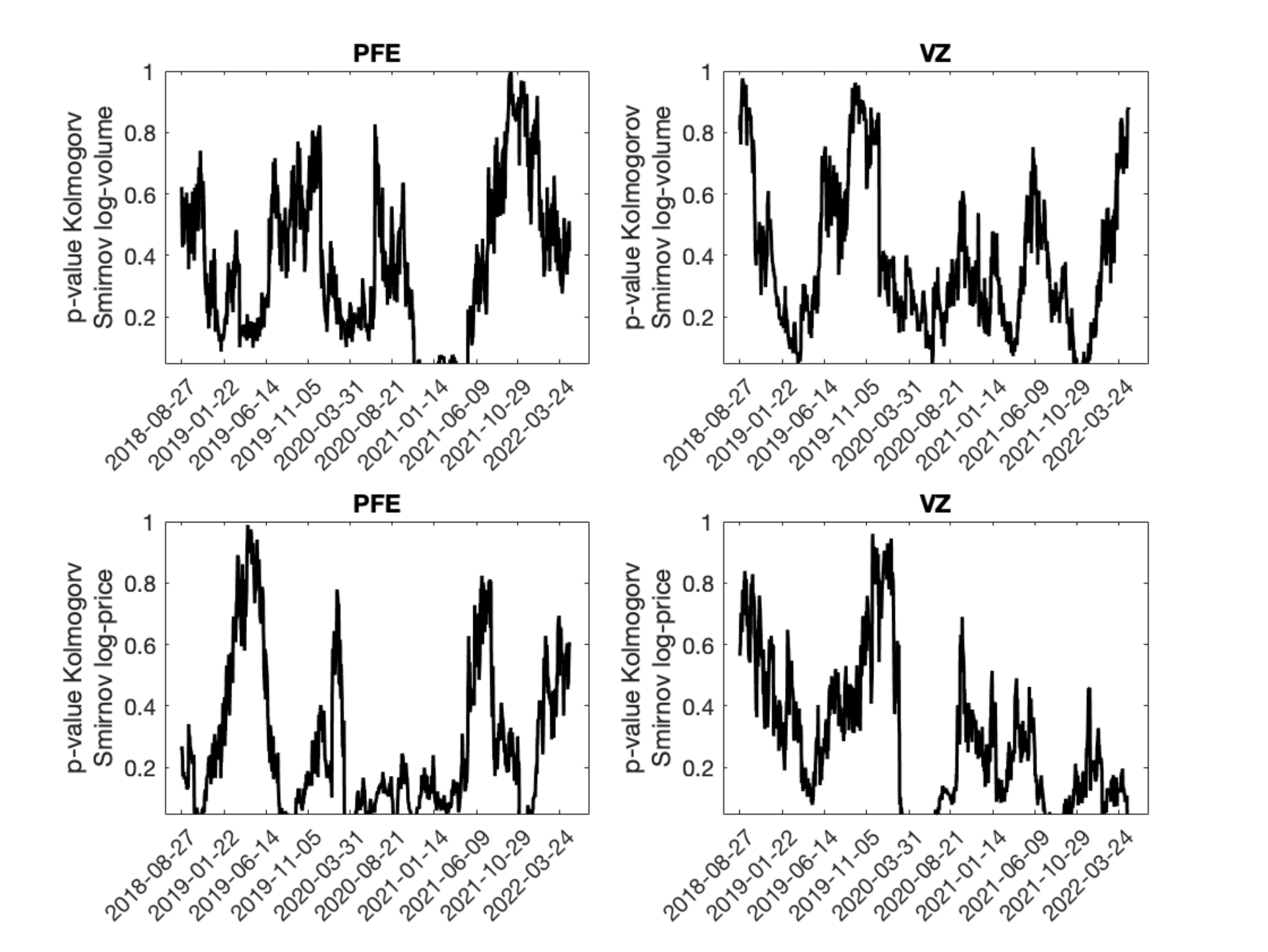}
 \caption{\small Kolmogorov-Smirnov test applied to the log-volume (upper panels) and log-price (lower panels) for PFE and VZ stocks, with the null hypothesis assuming normality of the time series. The $y$-axis range spans from $0.05$ to $1$.}\label{KS_Test_PFE_VZ}
\end{figure}
  \end{center}
\begin{center}
\begin{figure}[h!]
\centering\includegraphics[scale=0.32]{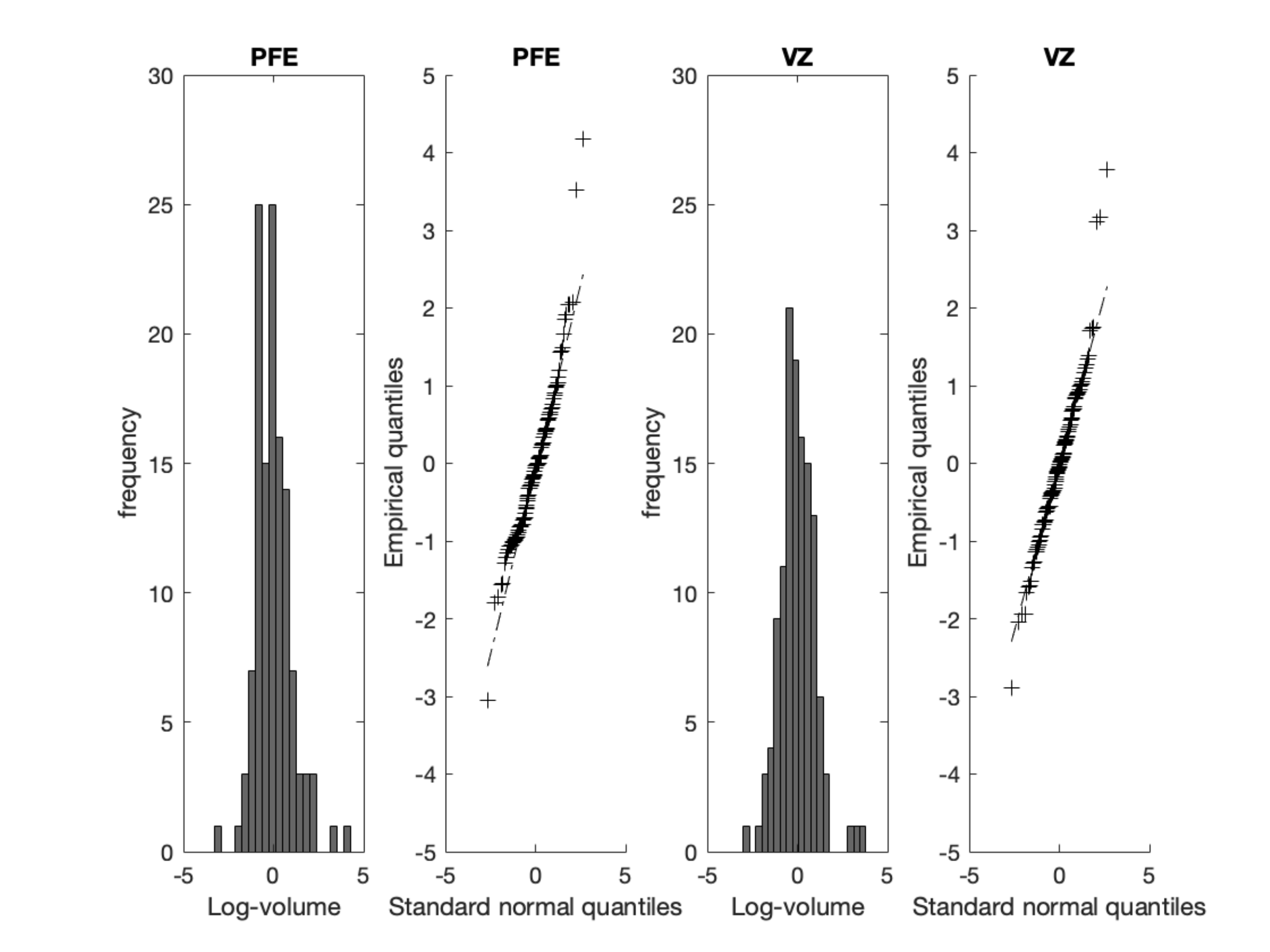}
 \caption{\small Frequency histogram and QQ-plot of the standardized daily log volumes  (from February 28th, 2018, to April 28th, 2022).}\label{Logvolume_PFE_VZ}
\end{figure}
  \end{center}
  \begin{figure}
\centering\includegraphics[scale=0.35]{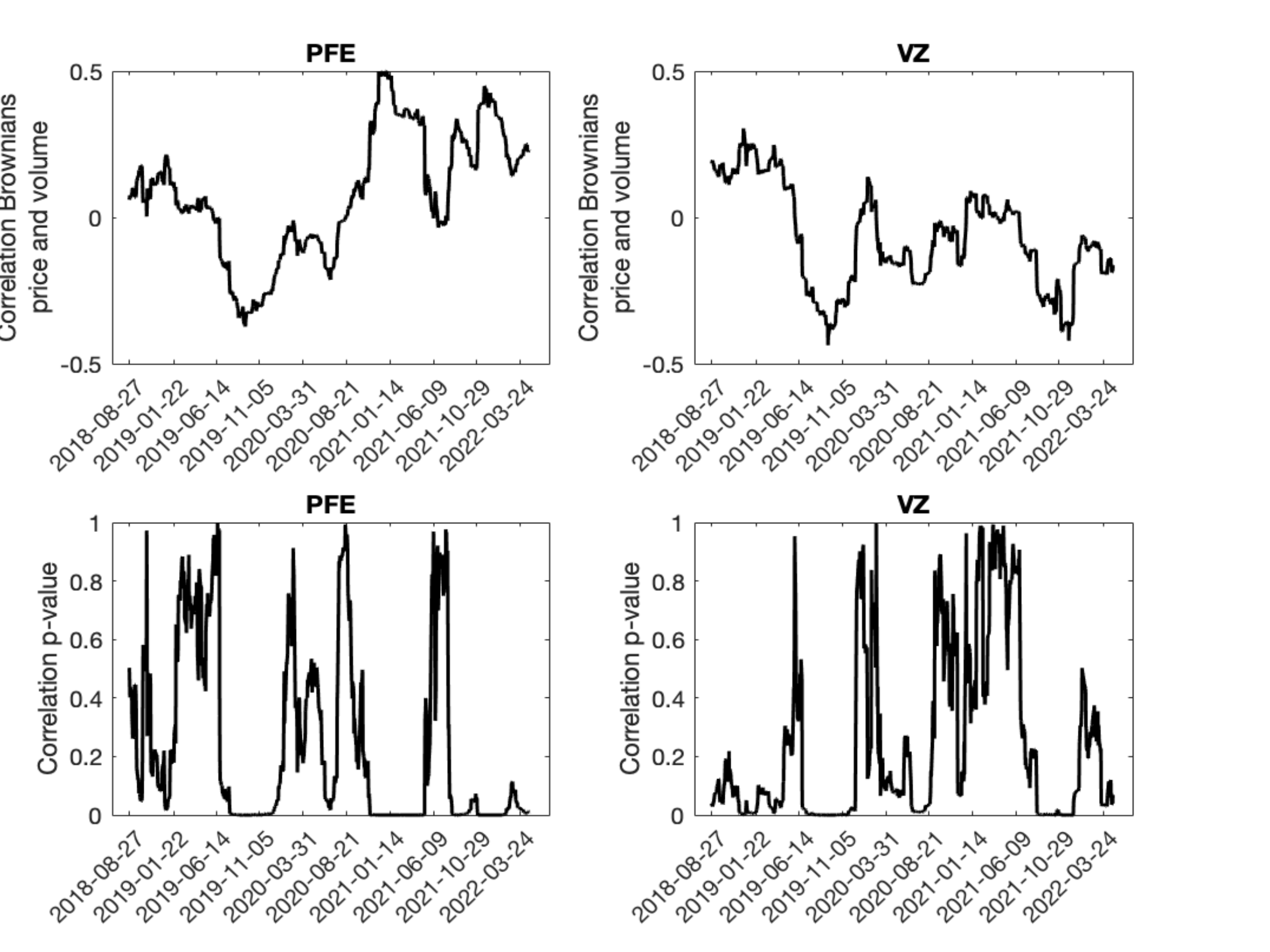}
\caption{\small Correlation coefficients between price and volume Brownian motions (top panels) and the corresponding p-values (bottom panels). }\label{correlation_PFE_VZ}
 \end{figure}
 
 Figure \ref{Logvolume_PFE_VZ} shows the frequency histogram and the QQ-plot of the standardized daily log-volumes relative to the time period considered. The results shown in Figure \ref{Logvolume_PFE_VZ} confirm that the observed total trading volume follows a GBM. 
 Figure \ref{correlation_PFE_VZ} shows the estimated Pearson correlation coefficients of the Brownian motions of the price and volume and their corresponding $p$-values.
Figure \ref{correlation_PFE_VZ} shows that we cannot reject the null hypothesis of zero correlation in several windows. This confirms that two different Brownian processes drive the dynamics of price and volume. 

Additionally, we assess the robustness of the results by employing the estimated noises and parameters to generate one-day-ahead price forecasts. In Figure \ref{Simula_PFE_VZ}, the true prices are represented by dashed bold lines, while the simulated prices are depicted by solid lines for both the PFE stock (left panel) and the VZ stock (right panel). The overlapping curves indicate the reliability of the estimated parameters.
\begin{figure}
\centering
   \includegraphics[scale=0.37]{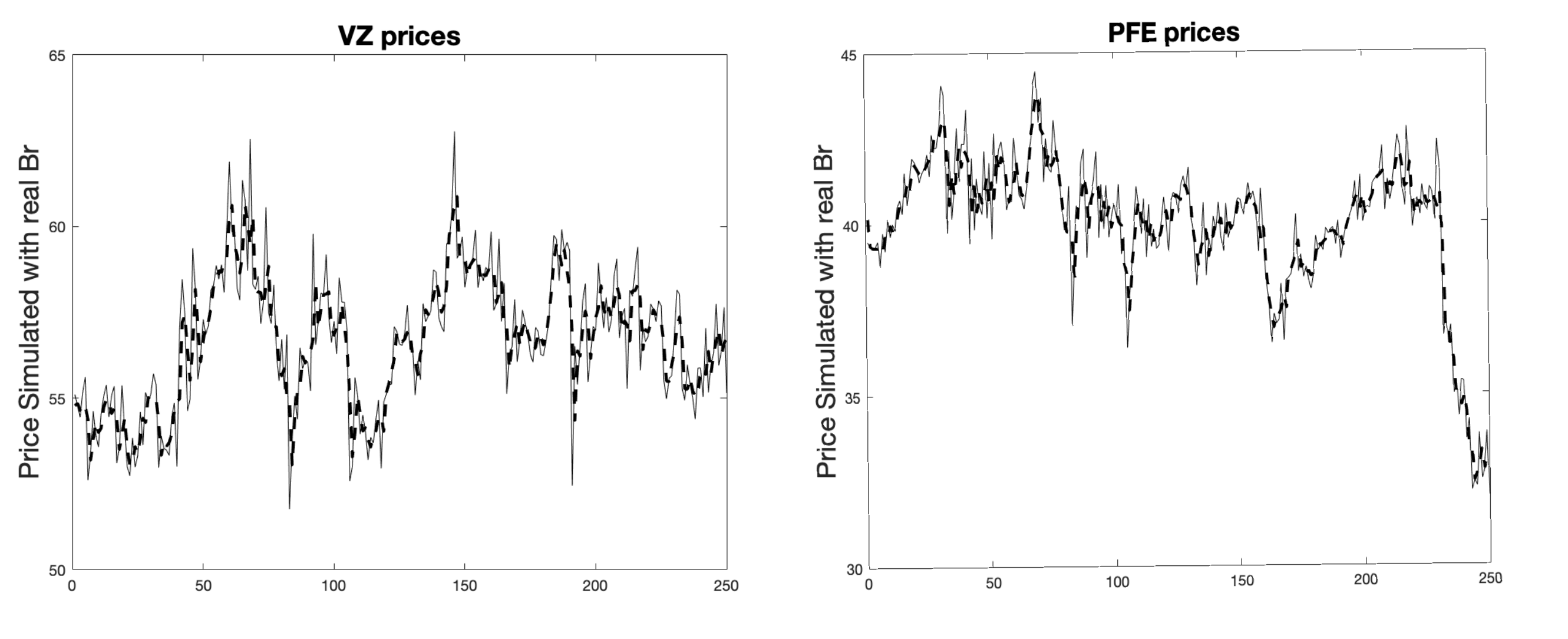}
\caption{One-day-ahead forecasts (solid line) and observed values (dashed bold line) for PFE (left panel) and VZ (right panel). }\label{Simula_PFE_VZ}
 \end{figure} 

Now, we show the relevant estimated quantities, that is, the drift-to-diffusion ratio, $\psi/\sigma$, of the volume time series compared to the theoretical ratio prescribed by the ORPMM model, $(\mu-r-\sigma^2)/\sigma$ (Fig. \ref{Ratio_PFE_VZ}), the volume and market price of risk (Fig. \ref{Risk_premium_PFE_VZ}) and the observed trading intensity (Fig. \ref{Trading_risk_PFE_VZ}).

Using the estimates  $\hat\mu$ and $\hat\sigma$ of $\mu$ and $\sigma$, respectively, we compute the theoretical ratio using ${\rm ratio}_{th}=(\hat\mu-r-\hat\sigma^2)/\hat\sigma$. 
Analogously, using estimates $\hat\mu^v\approx \eta\psi$ and $\hat\sigma^v\approx \eta\sigma$, we compute  the drift-to-diffusion ratio:
\begin{eqnarray}\label{obs_ratio}
 \textrm{\rm ratio}_{obs}=\frac{\hat\mu^v}{\hat\sigma^v}.
 \end{eqnarray}

\begin{figure}
\centering \includegraphics[scale=0.35]{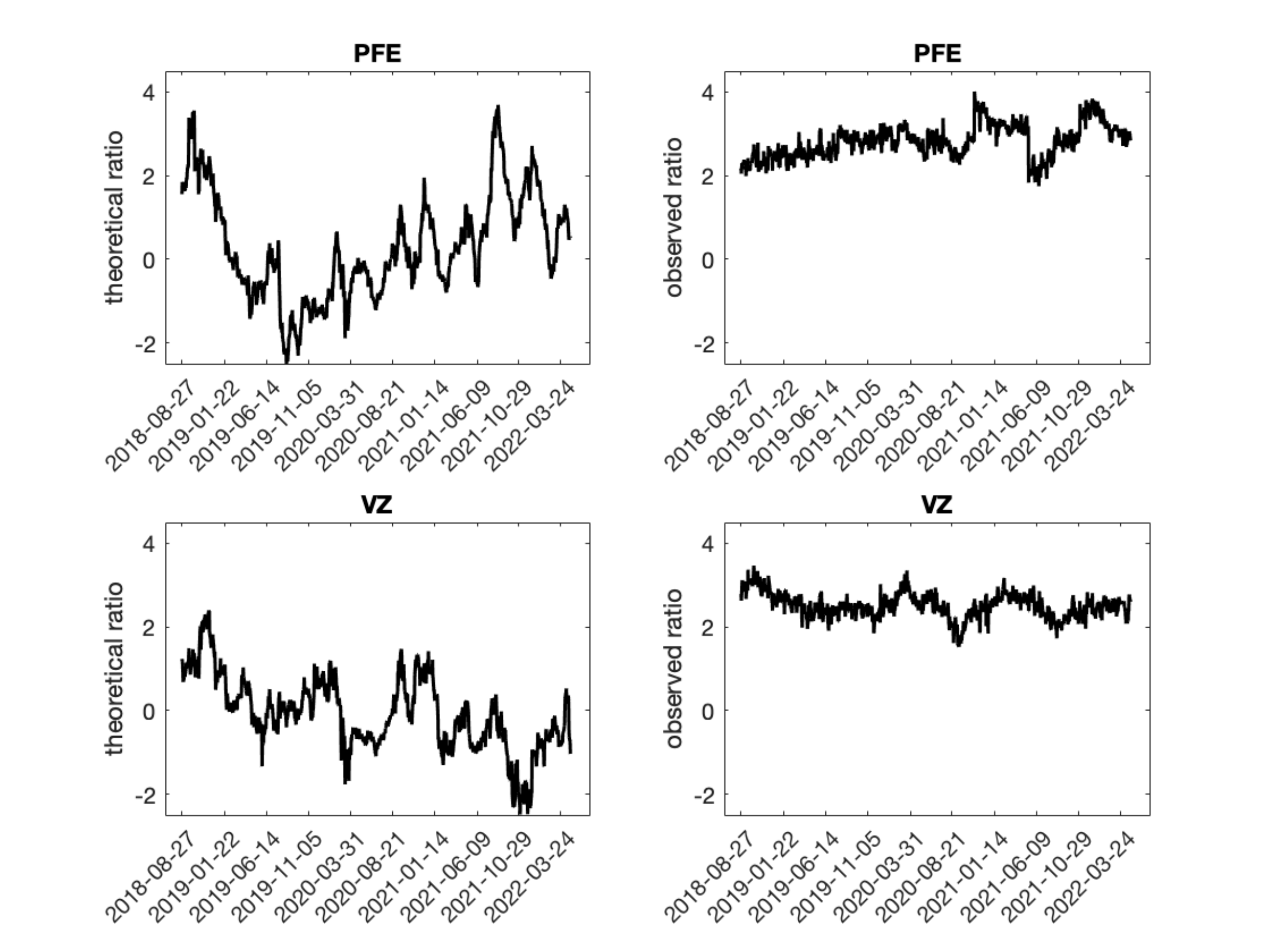}
\caption{\small  Comparison of ORPMM's theoretical drift-to-diffusion ratio of the volume dynamics with the observed values of the ratio. Linear regression with intercept only: PFE -- Intercept = 2.82, Standard deviation = 0.0132; VZ -- Intercept = 2.51, Standard deviation = 0.0099.
}\label{Ratio_PFE_VZ}
 \end{figure}
Upon examining Figure \ref{Ratio_PFE_VZ}, it is evident that the observed ratio does not coincide with the theoretical ORPMM ratio. More interestingly, the observed ratio is approximately constant in the time period considered and its range spans 2--4. This intriguing result appears to confirm the experiment outlined in \cite{KT1991}, who demonstrated that individuals tend to accept a bet over a certain monetary value only if the potential win is at least 2.5 times higher.
We check the value of the observed ratio by using a linear model with only intercept. The model yields the following results:  PFE intercept equal to 2.82 with standard deviation 0.0132;   VZ intercept equal to 2.51 with standard deviation 0.0099. These results are in line with the experiment of \cite{KT1991}.
\begin{figure}
\centering
\hspace{-1cm} \includegraphics[scale=0.25]{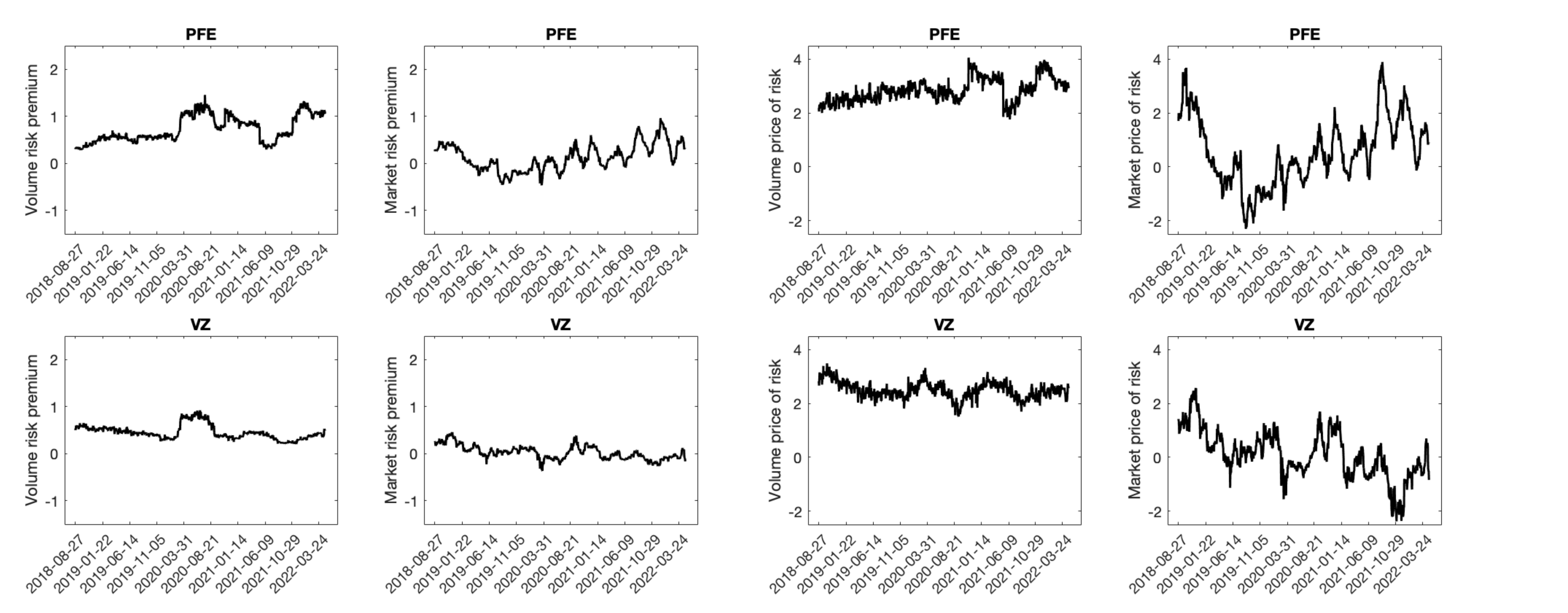}\caption{Volume risk premium, $\psi+\rho\sigma^2$ (first column, left panels), and the volume price of risk, $(\psi+\rho\sigma^2)/\sigma$ (first column, right panels). The market risk premium and the price of risk are in the second columns of the left and right panels, respectively. }\label{Risk_premium_PFE_VZ}
 \end{figure}

In the left panels of Figure \ref{Risk_premium_PFE_VZ} we compare the volume risk premium, 
$\psi+\rho\sigma^2$, (first column -- left panels) with the market risk premium, $\mu-r$ (second column -- left panels), while the right panels show the volume price of risk, 
$(\psi+\rho\sigma^2)/\sigma$ (first column -- right panels) and the market price of risk, 
$(\mu-r)/\sigma$ (first column -- right panels).

Remarkable, the volume price of risk remains relatively constant. Conducting linear regression with intercept only yields the following results: the PFE intercept equals 2.84 with a standard deviation of 0.014, while the VZ intercept equals 2.49 with a standard deviation of 0.01.

Regarding the market price of risk, we find an intercept of 0.549 with a standard deviation of 0.041 for PFE, and an intercept of 0.0914 with a standard deviation of 0.02 for VZ.

The volume price of risk can be easily interpreted from a behavioral standpoint. Market makers trading Pfizer stock, both pre-Covid and post-Covid, face slightly higher risks compared to those trading a more stable stock like Verizon Communications. Consequently, to participate in this market rather than sticking to bonds, they demand a premium per unit of risk of approximately three times the risk they face.

The interpretation of the market price of risk is straightforward from a comparative perspective, as the price for investing in PFE is approximately five times larger than investing in VZ. However, interpreting this in terms of a bet, as outlined by \cite{KT1991}, is challenging because a price of one half is meaningless in this framework.

Finally, we turn our attention to the ``ideal ORPMM world" scenario, where the discounted trading strategy reduces to a martingale with the trading price of risk (denoted as $TP$) equal to the expression (\ref{def_mkt}), i.e. $TP=\left(\mu-r+ (\psi+\rho\sigma^2)\eta\right)/(\sigma\sqrt{1+\eta^2+2\rho\eta})$. This trading price of risk incorporates both the price risk and the volume risk associated with the trading intensity $\eta$. Specifically, when the trading intensity is small, the price of risk reduces to the market price of risk. However, higher trading intensities may lead to an increase in the market price of risk, potentially exceeding the volume-adjusted price of risk.

\begin{figure}
\centering
\includegraphics[scale=0.25]{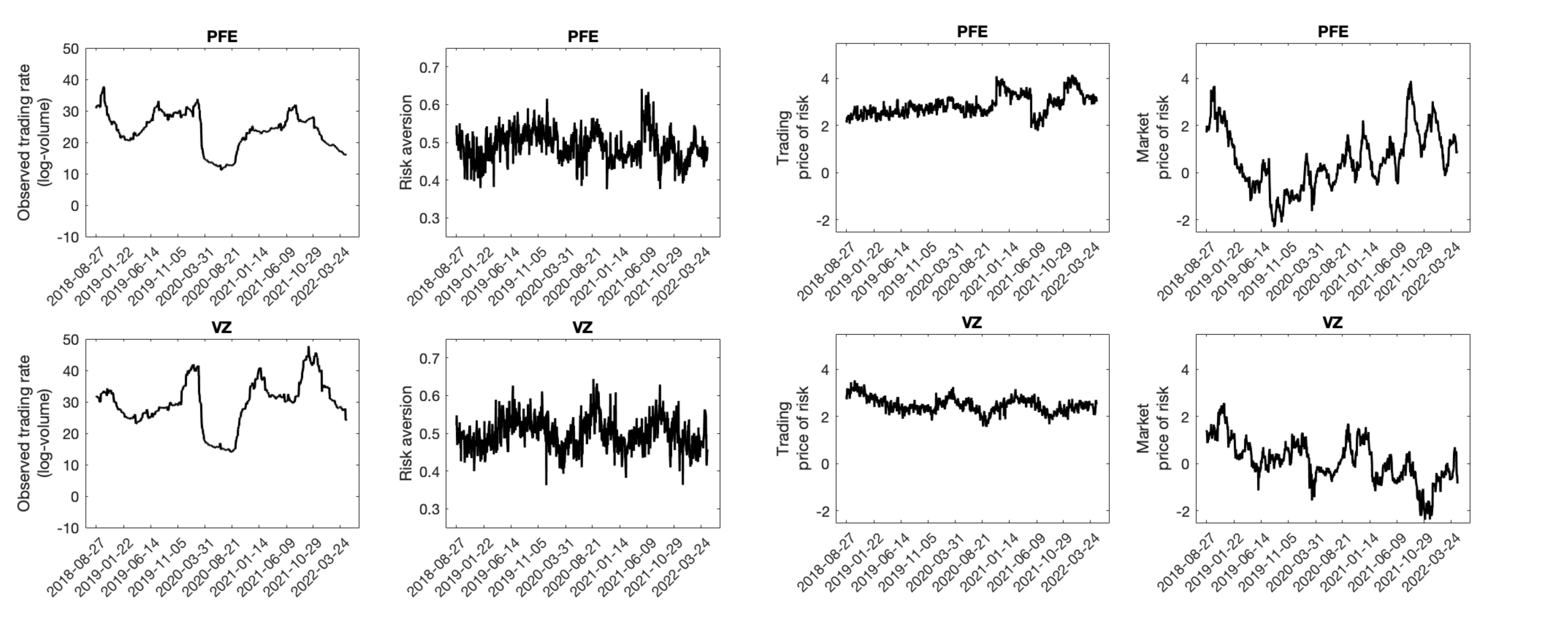}\caption{Trading intensity and risk aversion (left panels); trading price of risk and market price of risk (right panels).
 }\label{Trading_risk_PFE_VZ}
 \end{figure} 
 
In Figure \ref{Trading_risk_PFE_VZ}, the left panels show the estimated trading intensity $\eta$ and risk aversion $\bar{\gamma}$, while the right panels show the trading price of risk and the market price of risk. We conducted linear regressions with intercept-only models to estimate risk aversion and trading price of risk. We obtain the following results:\\
For the PFE stock: Risk aversion intercept = 0.491 (standard deviation = 0.00135), and the trading price of risk intercept = 2.869 (standard deviation = 0.0149).\\
For the VZ stock: Risk aversion intercept = 0.499 (standard deviation = 0.00139), and the trading price of risk intercept = 2.489 (standard deviation = 0.0105).

We observe that the trading price of risk closely resembles the volume price of risk and is consistent with the findings of \cite{KT1991}. It is remarkable how the risk aversion remains constant over time for market makers operating in these assets.

\subsection{Market Indices}\label{SPX_NA}

In this section we analyze two market indices, S\&P 500 (SPX) and NASDAQ (NA), from September 17th, 2018, to September 15th, 2023.  We repeat the experiment of Section \ref{PFE_VZ}.

Figure \ref{KS_Test_SPX_NA} displays the p-values (on the $y$-axis) obtained from the Kolmogorov-Smirnov test, where the null hypothesis assumes normality of the time series. This test is applied to both the log-volume (upper panels) and the log-price (lower panels) for S\&P 500 and NasdAQ within each window utilized in the estimation process. The $y$-axis  ranges from 0.05 to 1.

We observe that the p-values for the log-volume are generally greater than or equal to 0.05 for several dates, except during the Covid period. However, the volume of the SPX index exhibits a worse behavior compared to that of the NA index. Regarding the prices, normality is particularly rejected during the Covid years (2020-2021) for both the SPX and NA prices.

\begin{figure}[h!]
\centering\includegraphics[scale=0.35]{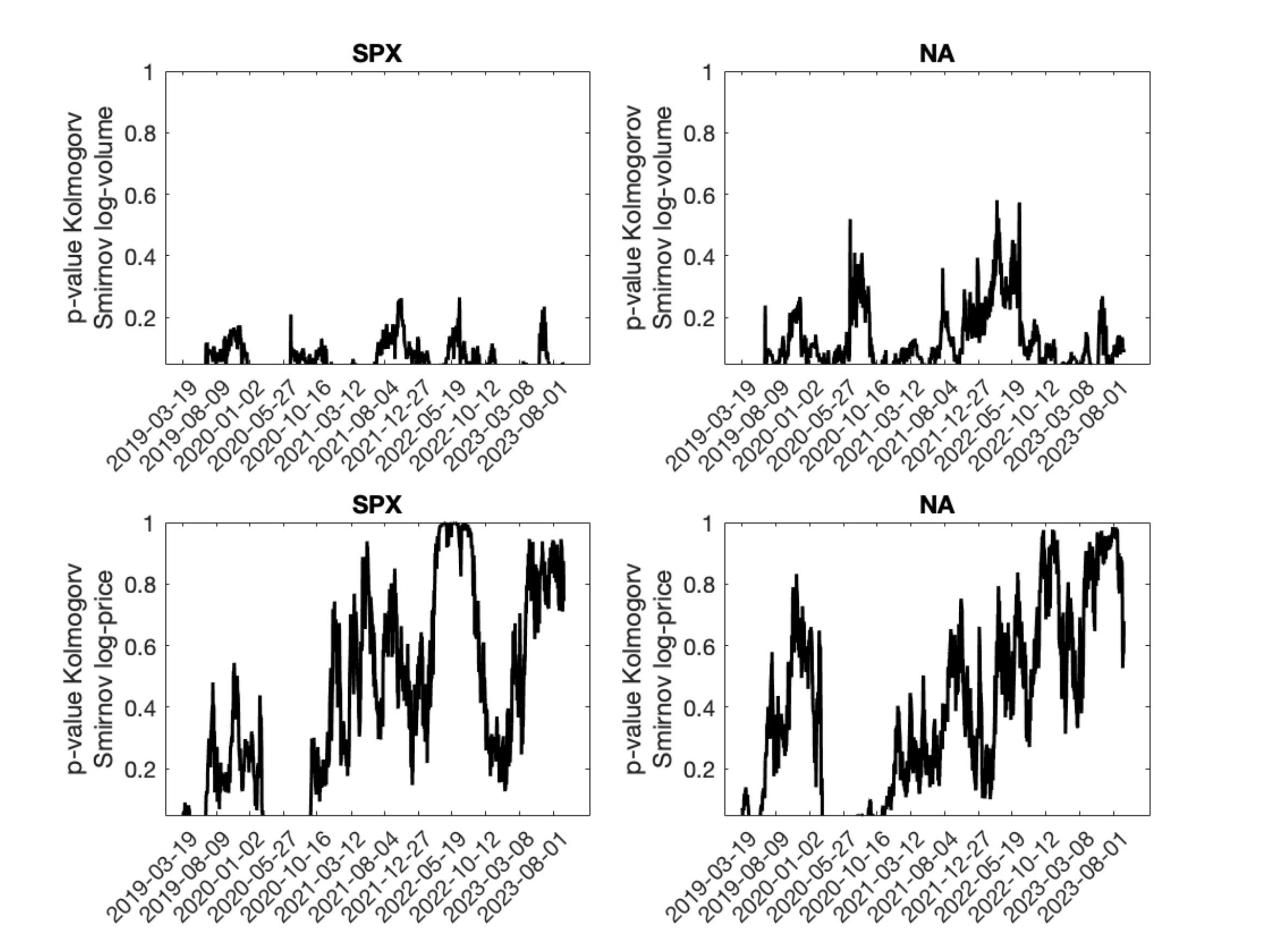}
 \caption{\small Kolmogorov-Smirnov test applied to the log-volume (upper panels) and log-price (lower panels) for SPX and NA indices, with the null hypothesis assuming normality of the time series. The $y$-axis range spans from $0.05$ to $1$.}\label{KS_Test_SPX_NA}
\end{figure}
Figure \ref{Logvolume_SPX_NA} shows the  frequency histogram and the QQ-plot of the standardized daily log-volumes relative to the time period considered. These results confirm that the observed total trading volume follows a GBM. 
\begin{figure}[h!]
\centering \includegraphics[scale=0.32]{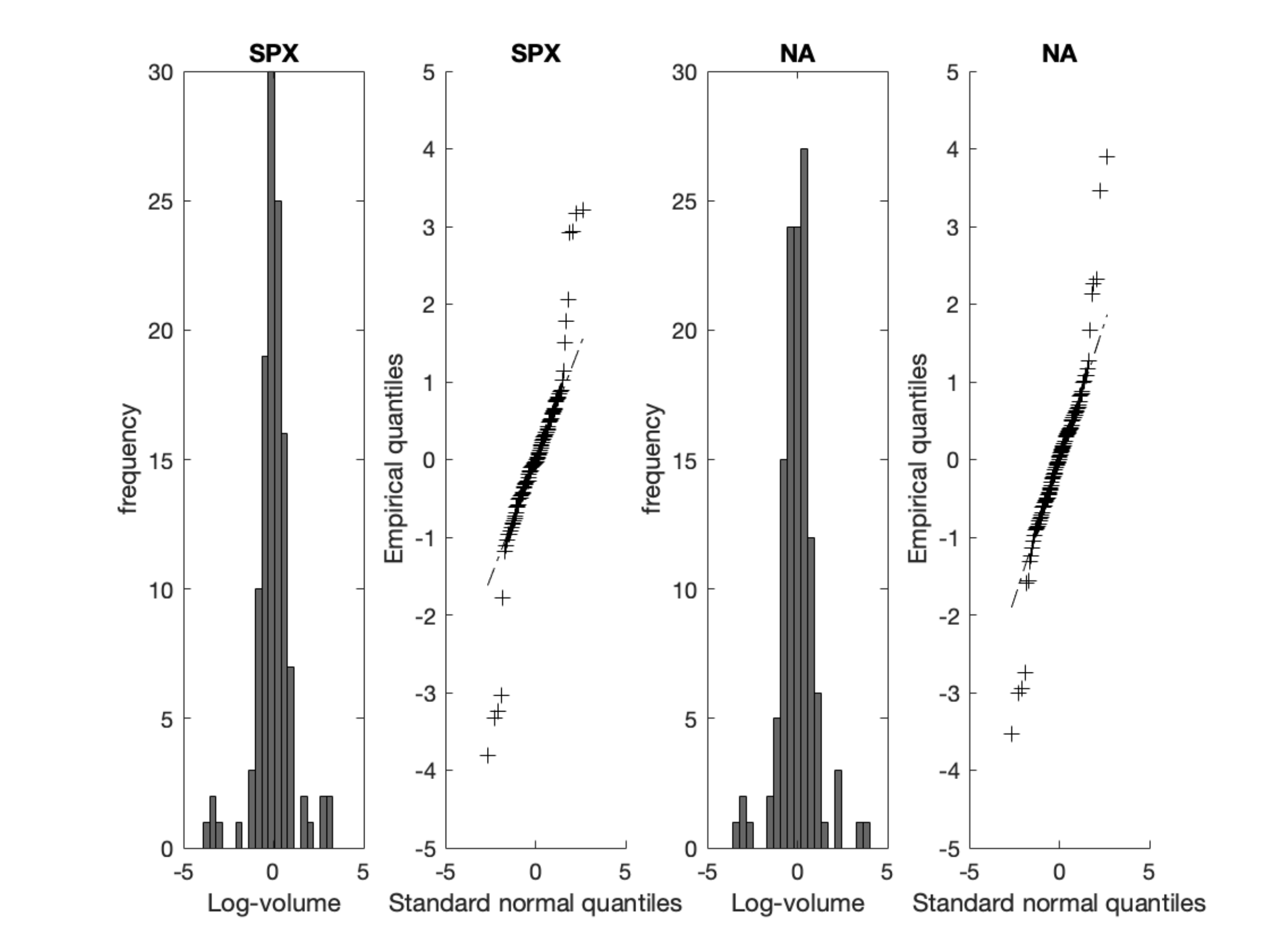}
 \caption{\small Frequency histogram and QQ-plot of the standardized daily log volumes  (from September 17th, 2018, to September 15th, 2023).}\label{Logvolume_SPX_NA}
\end{figure}
 Figure \ref{correlation_SPX_NA} shows the estimated Pearson correlation coefficients of the Brownian motions of the price and volume and their corresponding $p-value$.
\begin{figure}
\centering\includegraphics[scale=0.35]{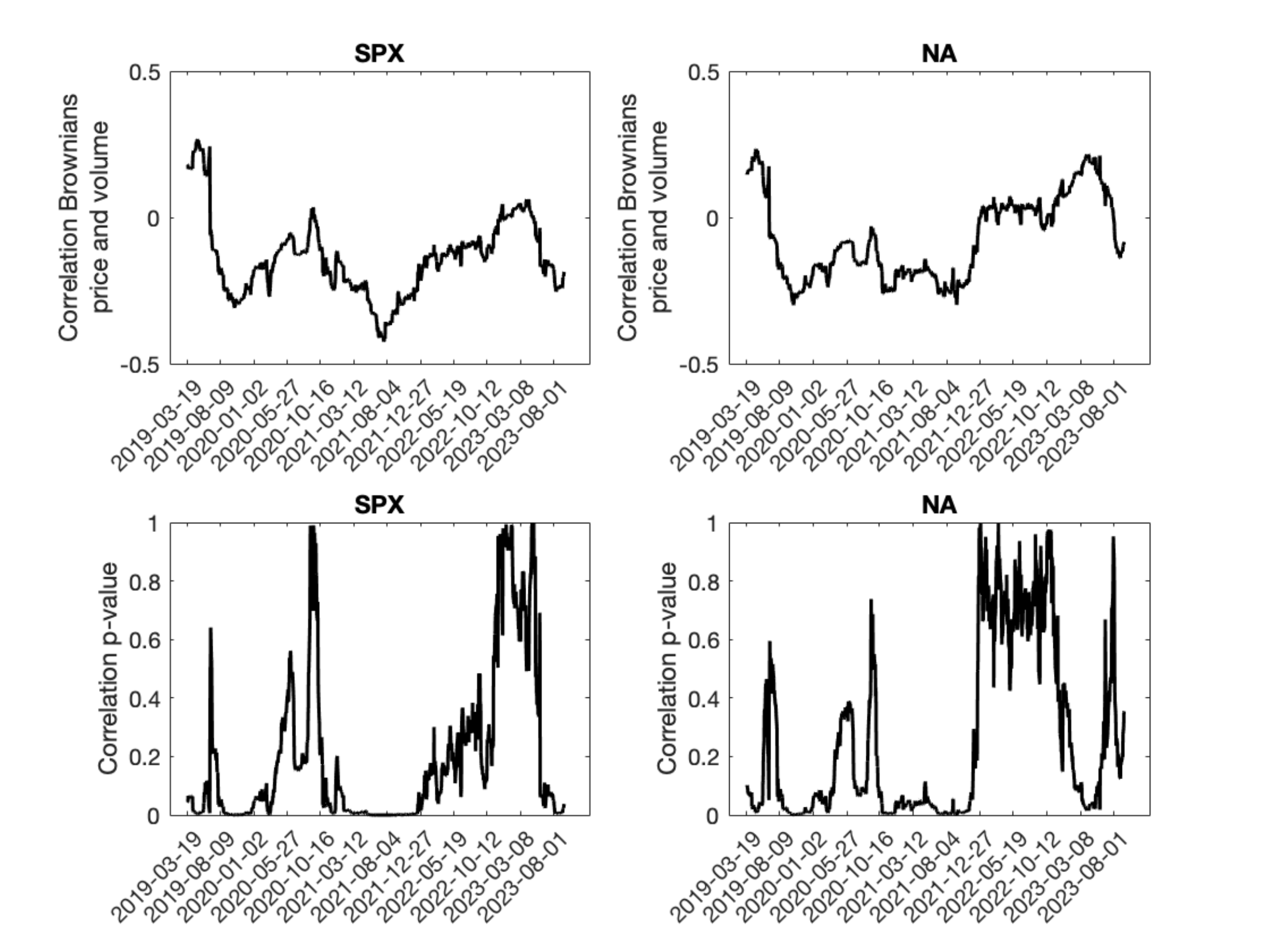}
\caption{\small Correlation coefficients between price and volume Brownian motions (top panels) and the corresponding p-values (bottom panels). }\label{correlation_SPX_NA}
 \end{figure}

Figure \ref{correlation_SPX_NA} indicates that we cannot reject the null hypothesis of zero correlation in several windows. This suggests that two distinct Brownian motions drive the dynamics of price and volume.

Additionally, we assess the robustness of the results by employing the estimated noises and parameters to generate one-day-ahead price forecasts. In Figure \ref{Simula_SPX_NA}, the true prices are represented by dashed bold lines, while the simulated values are indicated by solid lines for both the SPX index (left panel) and the NA index (right panel). The overlapping curves indicate the reliability of the estimated parameters.
\begin{figure}
\centering\includegraphics[scale=0.23]{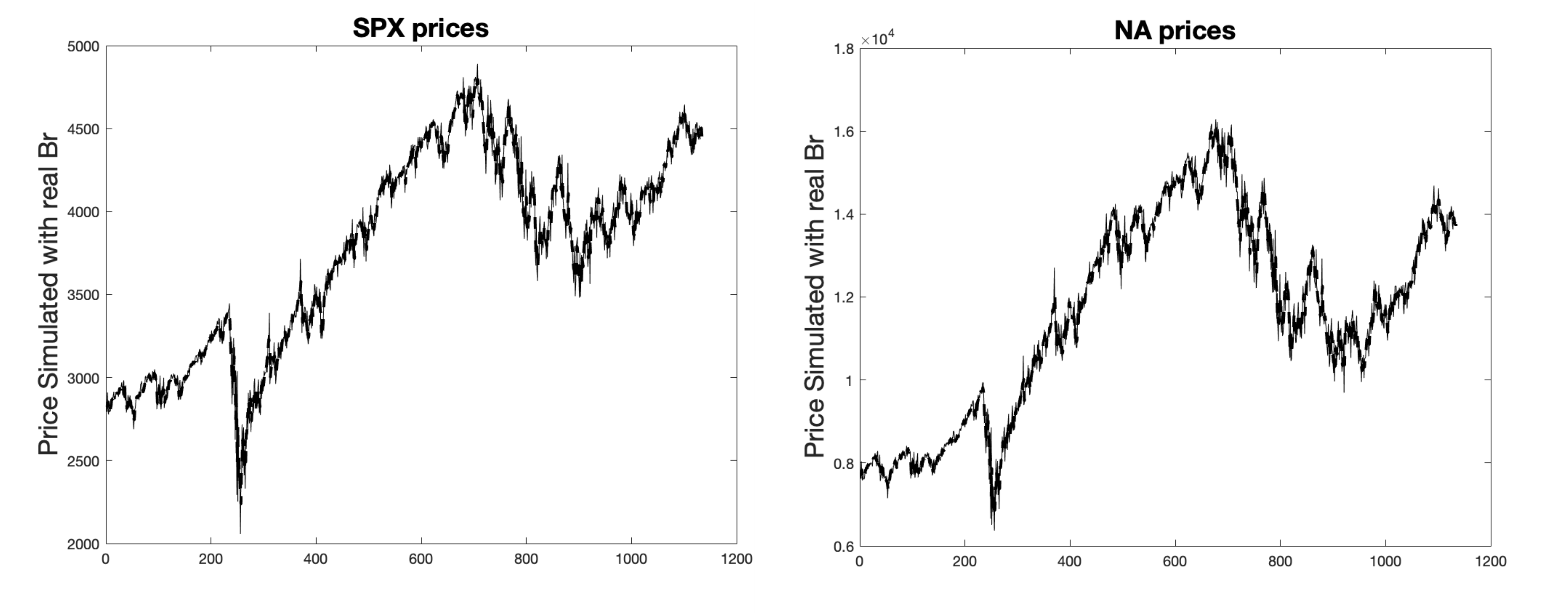}\caption{One-day-ahead forecasts (solid line) and observed values (dashed bold line) for SPX (left panel) and NA (right panel). }\label{Simula_SPX_NA}
 \end{figure} 

Next, we analyze the estimates of the drift-to-diffusion ratio, $\psi/\sigma$, and compare them with the ratio prescribed by ORPMM model, $(\mu-r-\sigma^2)/\sigma$ (see Fig. \ref{Ratio_SPX_NA}). Additionally, we examine the volume price of risk and the market price of risk (Fig. \ref{Risk_premium_SPX_NA}), along with the observed trading intensity (Fig. \ref{Trading_risk_SPX_NA}).

\begin{figure}
\centering\includegraphics[scale=0.35]{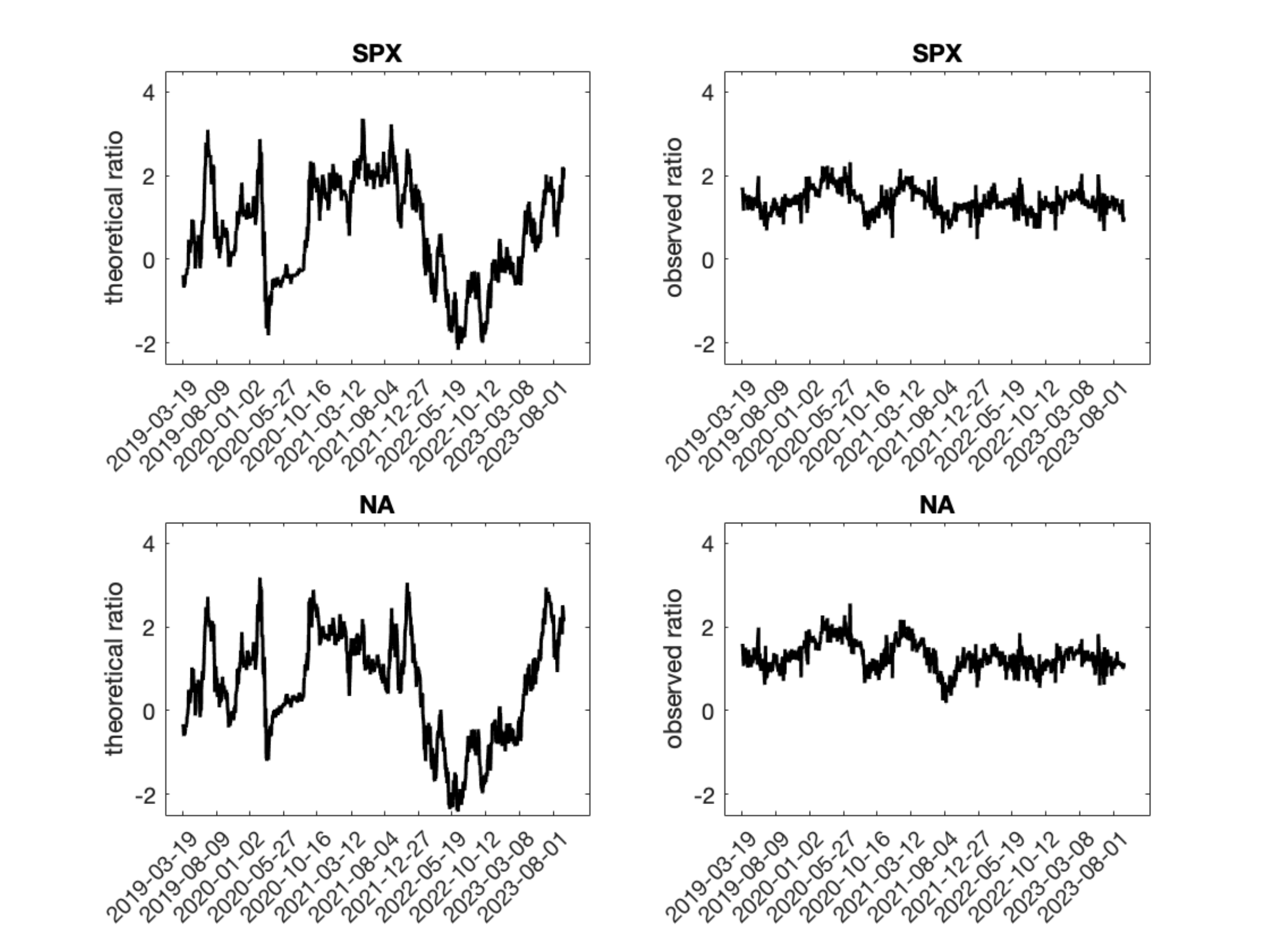}
\caption{\small  Comparison of the ORPMM theoretical drift-to-diffusion ratio of the volume dynamics with observed values of the ratio. Linear regression with intercept only: SPX -- Intercept = 1.38, Standard deviation = 0.00845; NA -- Intercept = 1.28, Standard deviation = 0.0102.
}\label{Ratio_SPX_NA}
 \end{figure}

Upon examining Figure \ref{Ratio_SPX_NA}, it is evident that the observed ratio does not coincide with the ORPMM theoretical ratio. More interestingly, the observed ratio remains approximately constant throughout the considered time period.

We further assess the value of the observed ratio by conducting a linear regression with intercept only, resulting in the following outcomes: SPX intercept equals 1.38 with a standard deviation of 0.00845, and NA intercept equals 1.28 with a standard deviation of 0.0102.

These results are consistent with the fact that PFE and VZ stocks are riskier than the SPX and NA indices. Specifically, the ratios of the indices are half those of the stocks.

In the left panels of Figure \ref{Risk_premium_SPX_NA} we compare the volume risk premium, 
$\psi+\rho\sigma^2$, (first column -- left panels) with the market risk premium, $\mu-r$ (second column -- left panels), while the right panels show the volume price of risk, 
$(\psi+\rho\sigma^2)/\sigma$ (first column -- right panels) and the market price of risk, 
$(\mu-r)/\sigma$ (first column -- right panels).

\begin{figure}
\centering
\hspace{-1.2cm}\includegraphics[scale=0.25]{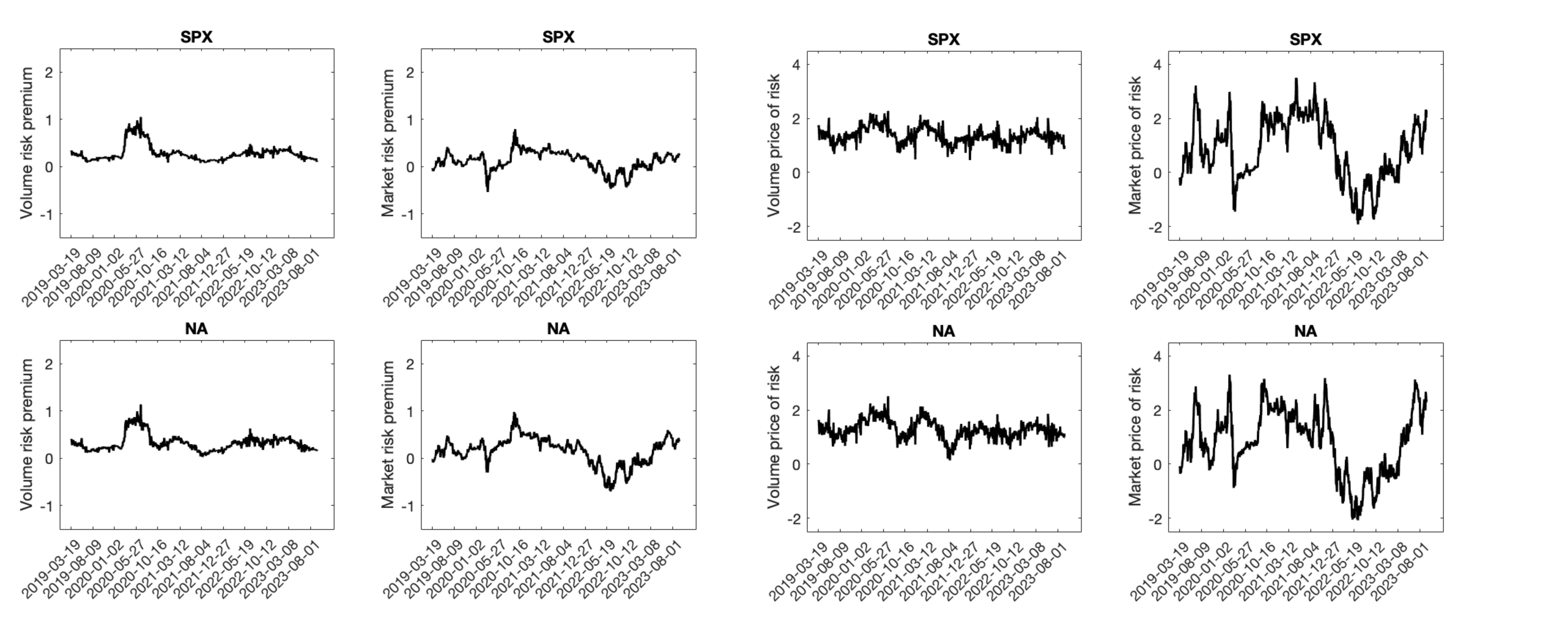}\caption{Volume risk premium, $\psi+\rho\sigma^2$ (first column, left panels), and the volume price of risk, $(\psi+\rho\sigma^2)/\sigma$ (first column, right panels). The market risk premium and the price of risk are in the second columns of the left and right panels, respectively. }\label{Risk_premium_SPX_NA}
 \end{figure} 

Notably, the volume price of risk remains relatively constant. Conducting linear regression with intercept only yields the following results: SPX intercept equals 1.36 with a standard deviation of 0.0084, while NA intercept equals 1.27 with a standard deviation of 0.01.

Regarding the market price of risk, we find an intercept of 0.844 with a standard deviation of 0.034 for PFE, and an intercept of 0.807 with a standard deviation of 0.035 for NA.

The volume price of risk indicates that market makers trading Pfizer or Verizon stocks face slightly higher risks compared to those trading SPX and NA indices. By comparing the volume price of risk, we can conclude that marker makers participating in the stock market rather than sticking to market indices demand a premium per unit of risk of approximately twice as high.

In contrast, the market price of risk for the indices remains substantially the same and comparable to those of the considered stocks.

Finally, we turn our attention to the trading price of risk in the ``ideal ORPMM world".
\begin{figure}
\centering
\includegraphics[scale=0.23]{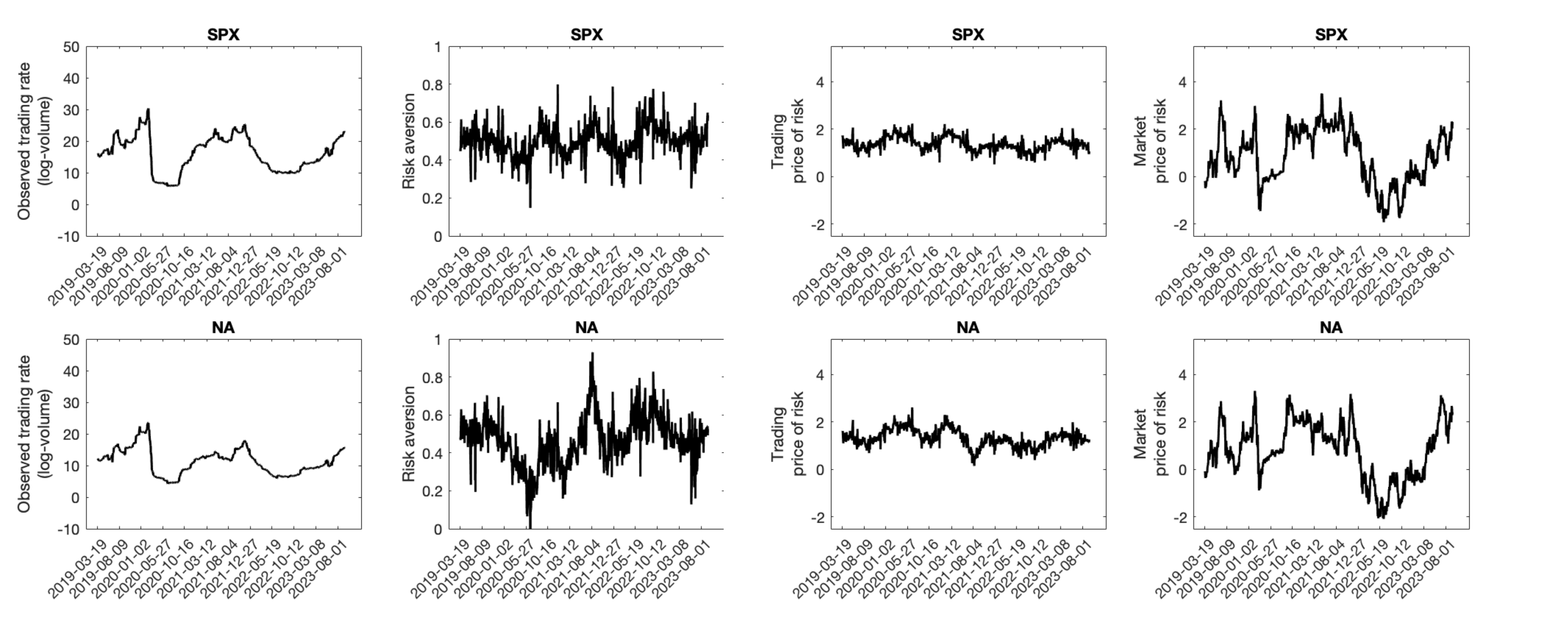}\caption{Trading intensity and risk aversion (left panels); trading price of risk and market price of risk (right panels).
 }\label{Trading_risk_SPX_NA}
 \end{figure}

In Figure \ref{Trading_risk_SPX_NA}, the left panels show the estimated trading intensity $\eta$ and risk aversion $\bar{\gamma}$, while the right panels show the trading price of risk and market price of risk. We conducted linear regressions with intercept-only models for risk aversion and trading price of risk, obtaining the following results:\\
For the SPX index: Risk aversion intercept = 0.499 (standard deviation = 0.0023), and trading price of risk intercept = 1.3826 (standard deviation = 0.0081).\\
For the NA index: Risk aversion intercept = 0.466 (standard deviation = 0.00384), and trading price of risk intercept = 1.3211 (standard deviation = 0.0107).

We observe that the trading price of risk closely resembles the volume price of risk and is consistent with the findings of  \cite{KT1991}. It is remarkable how the risk aversion remains constant over time for market makers operating in these assets.

\subsection{Exchange rates: BTC-USD and EUR-USD}\label{BTC_EUR}

In this section we analyze two exchange rates, Bitcoin to US Dollar (BTC-USD or BTC for short) and Euro to US-Dollar (EUR-USD or EUR for short), from September 30th, 2018, to September 23rd, 2023.  Also in this case we repeat the experiment of Section \ref{PFE_VZ}.

Figure \ref{KS_Test_BTC_EUR} displays the p-values (on the $y$-axis) obtained from the Kolmogorov-Smirnov test, where the null hypothesis assumes normality of the time series. This test is applied to both the log-volume (upper panels) and the log-exchange rate (lower panels) for BTC-USD  and EUR-USD within each window used in the estimation process. As in the previous sections, the $y$-axis  ranges from 0.05 to 1.

We observe that the p-values for the log-volume and log-rate of BTC-EUR are generally greater than or equal to 0.05 for several dates, except during the post Covid period (2021). The EUR-USD rate rejects the model for both volume and rate.  Thus, the results regarding the EUR-USD exchange rate could be not reliable.

\begin{figure}[h!]
\centering \includegraphics[scale=0.35]{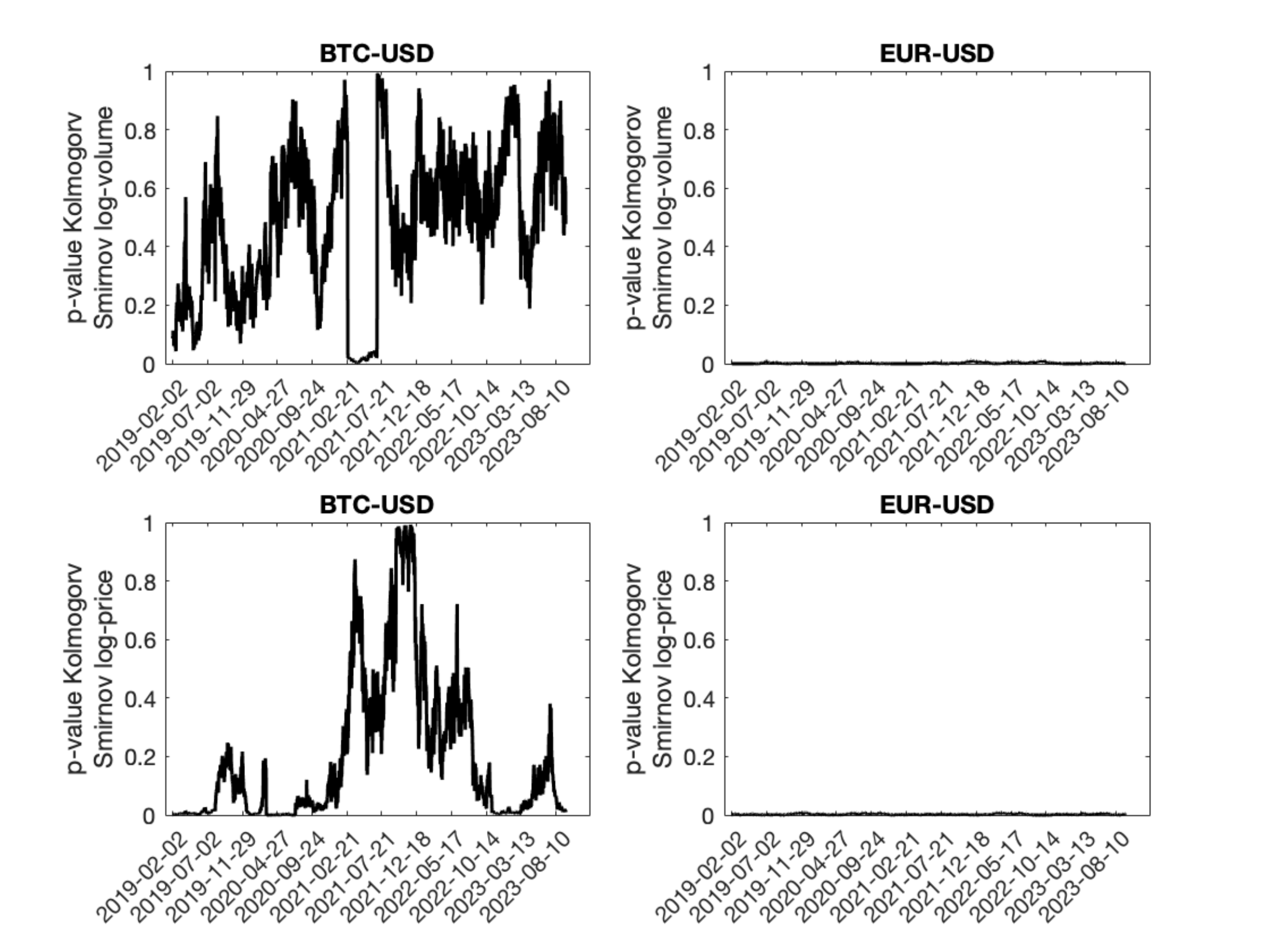}
 \caption{\small Kolmogorov-Smirnov test applied to the log-volume (upper panels) and log-exchange rate (lower panels) for BTC and EUR rates, with the null hypothesis assuming normality of the time series. The $y$-axis range spans from $0.05$ to $1$. }\label{KS_Test_BTC_EUR}
\end{figure}
Figure \ref{Logvolume_BTC_EUR} shows the  frequency histogram and the QQ-plot of the standardized daily log-volumes relative to the time period considered and it confirms that  the observed total trading volume follows a GBM. 
\begin{figure}[h!]
\centering \includegraphics[scale=0.32]{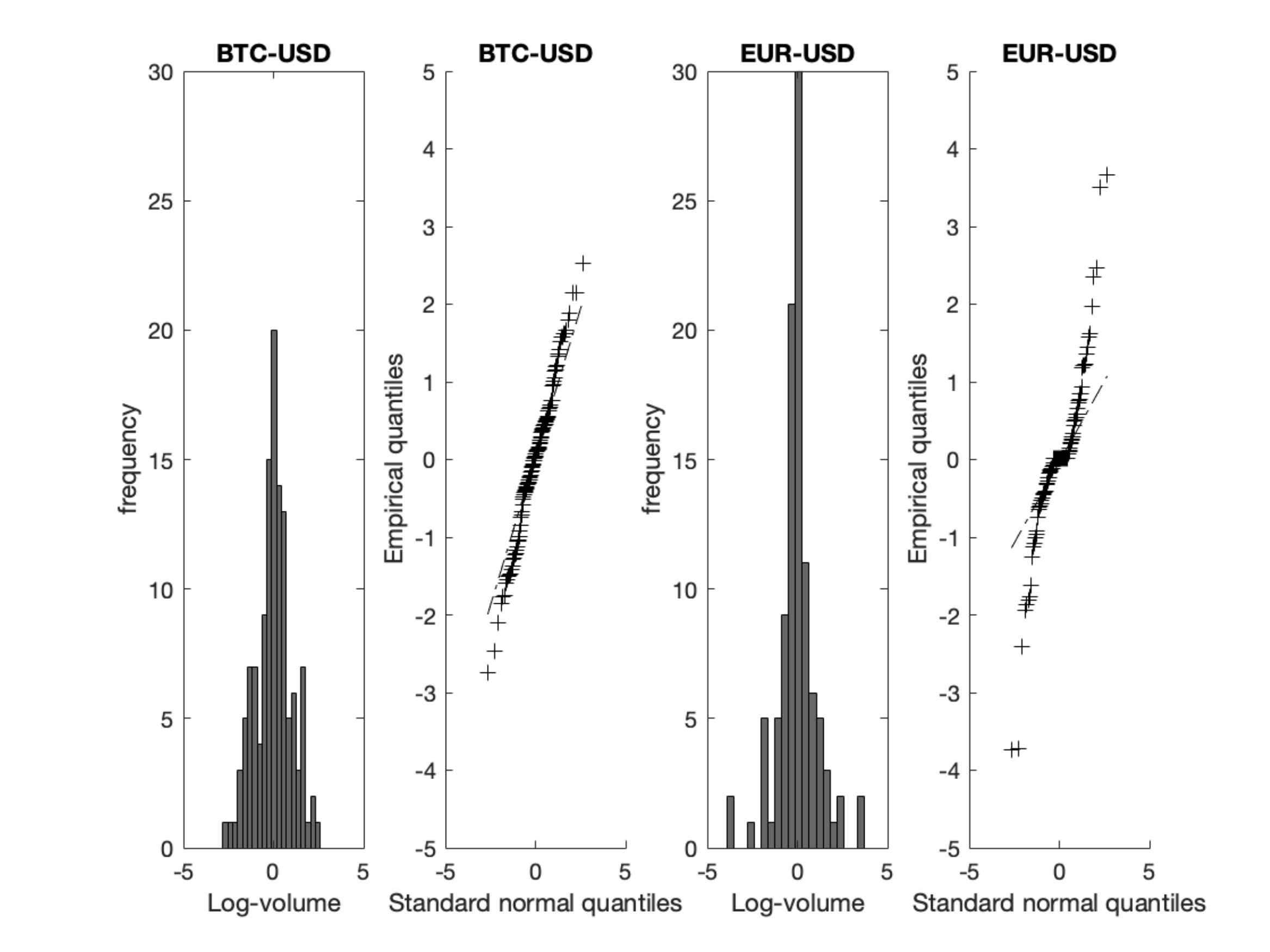}
 \caption{\small Frequency histogram and QQ-plot of the standardized daily log volumes  (from September 30th, 2018, to September 23rd, 2023).}\label{Logvolume_BTC_EUR}
\end{figure}
  
 Figure \ref{correlation_BTC_EUR} shows the estimated Pearson correlation coefficients of price and volume Brownian motions and the corresponding p-values.

\begin{figure}
\centering \includegraphics[scale=0.35]{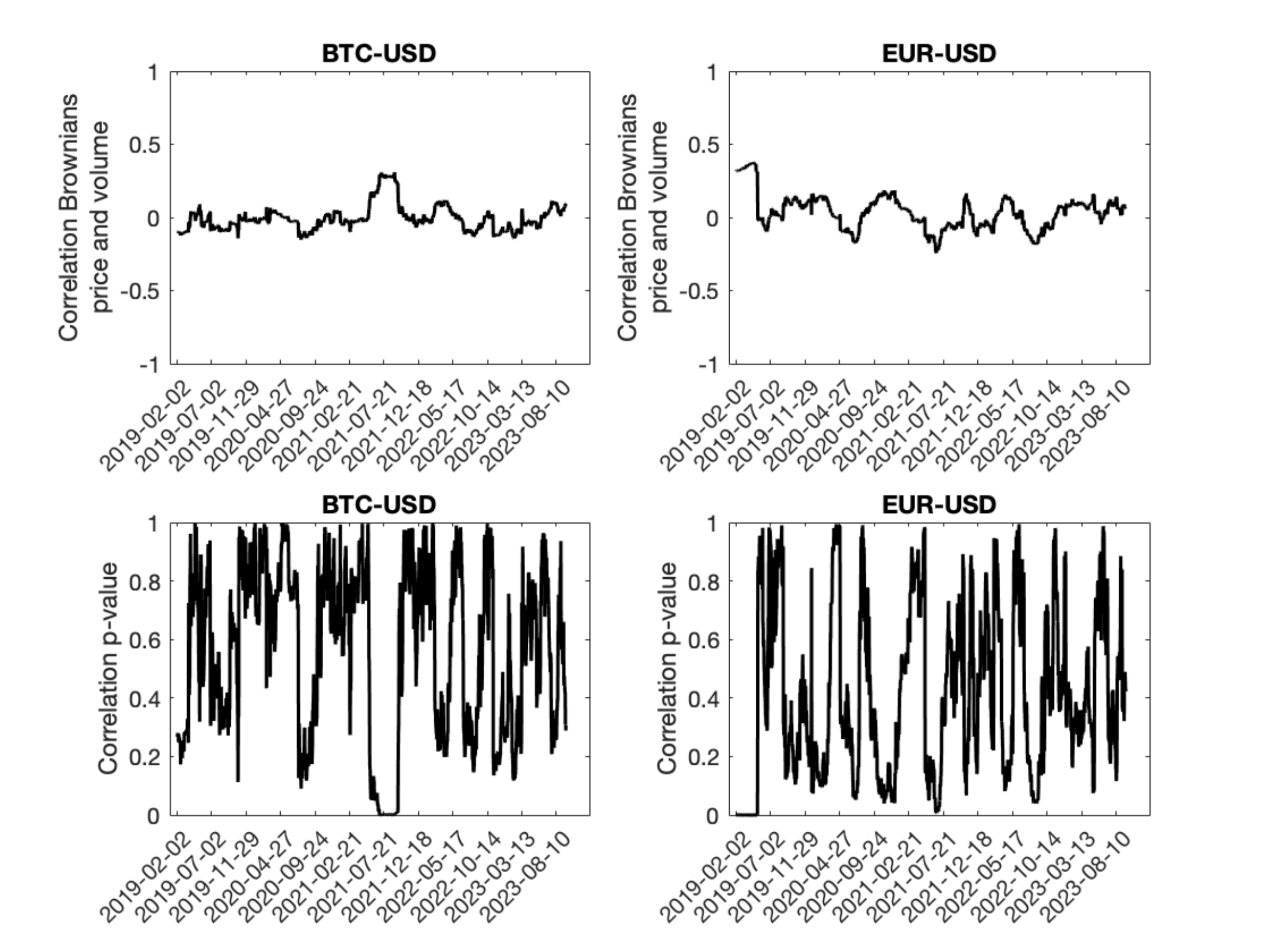}
\caption{\small Correlation coefficients between price and volume Brownian motions (top panels) and the corresponding p-values (bottom panels). }\label{correlation_BTC_EUR}
 \end{figure}

Figure \ref{correlation_BTC_EUR} indicates that we cannot reject the null hypothesis of zero correlation in several windows. This suggests that two distinct Brownian motions drive the dynamics of price and volume.

As in the previous sections, we perform an analysis of the robustness by employing the estimated noises and parameters to generate one-day-ahead price forecasts. In Figure \ref{Simula_BTC_EUR}, the true exchange rates are represented by dashed bold lines, while the simulated values are indicated by solid lines for both BTC-USD (left panel) and  EUR-USD index (right panel). The overlapping curves indicate the reliability of the estimated parameters.

\begin{figure}
\centering
\hspace{-1.2cm} \includegraphics[scale=0.23]{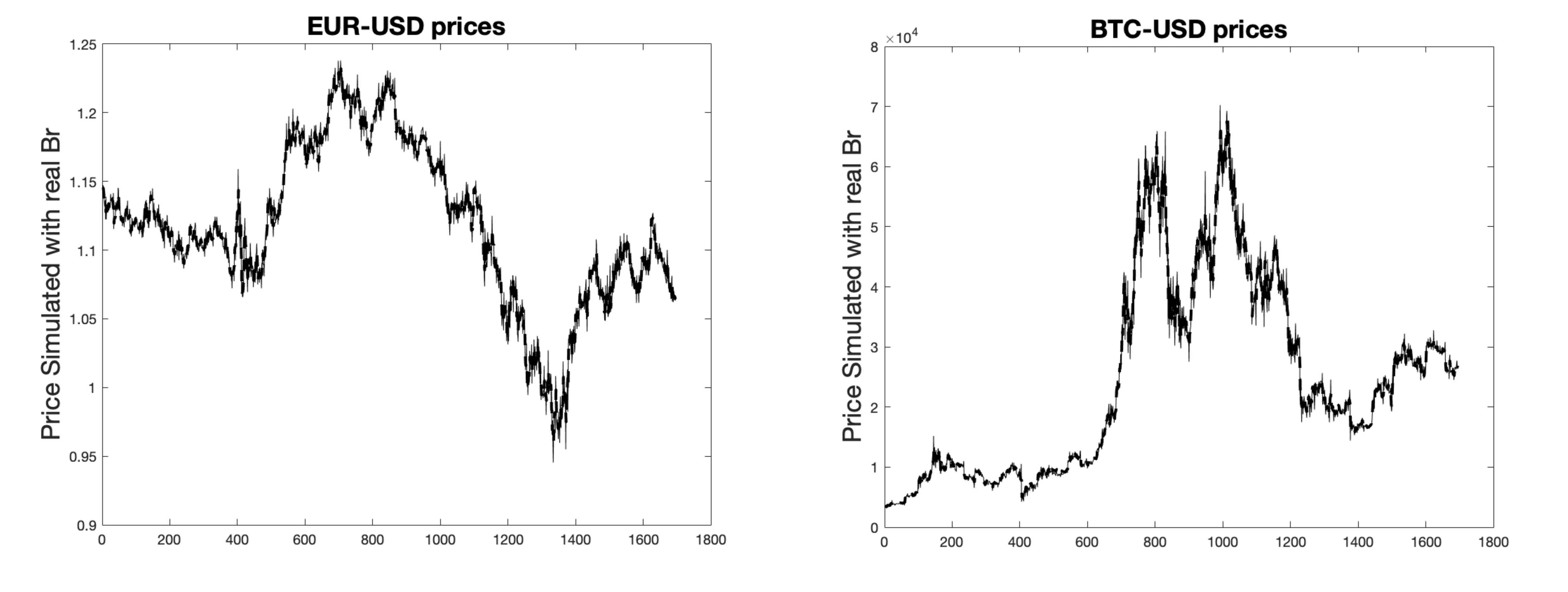}\caption{One-day-ahead forecasts (solid line) and observed values (dashed bold line) for BTC-USD (left panel) and EUR-USD (right panel). }\label{Simula_BTC_EUR}
 \end{figure}

Next, we analyze the estimates of the drift-to-diffusion ratio,  $\psi/\sigma$, and compare them with the ratio prescribed by ORPMM model, $(\mu-r-\sigma^2)/\sigma$ (see Fig. \ref{Ratio_BTC_EUR}). Additionally, we examine the volume price of risk and the market price of risk (Fig. \ref{Risk_premium_BTC_EUR}), along with the observed trading intensity (Fig. \ref{Trading_risk_BTC_EUR}).

\begin{figure}
\centering \includegraphics[scale=0.35]{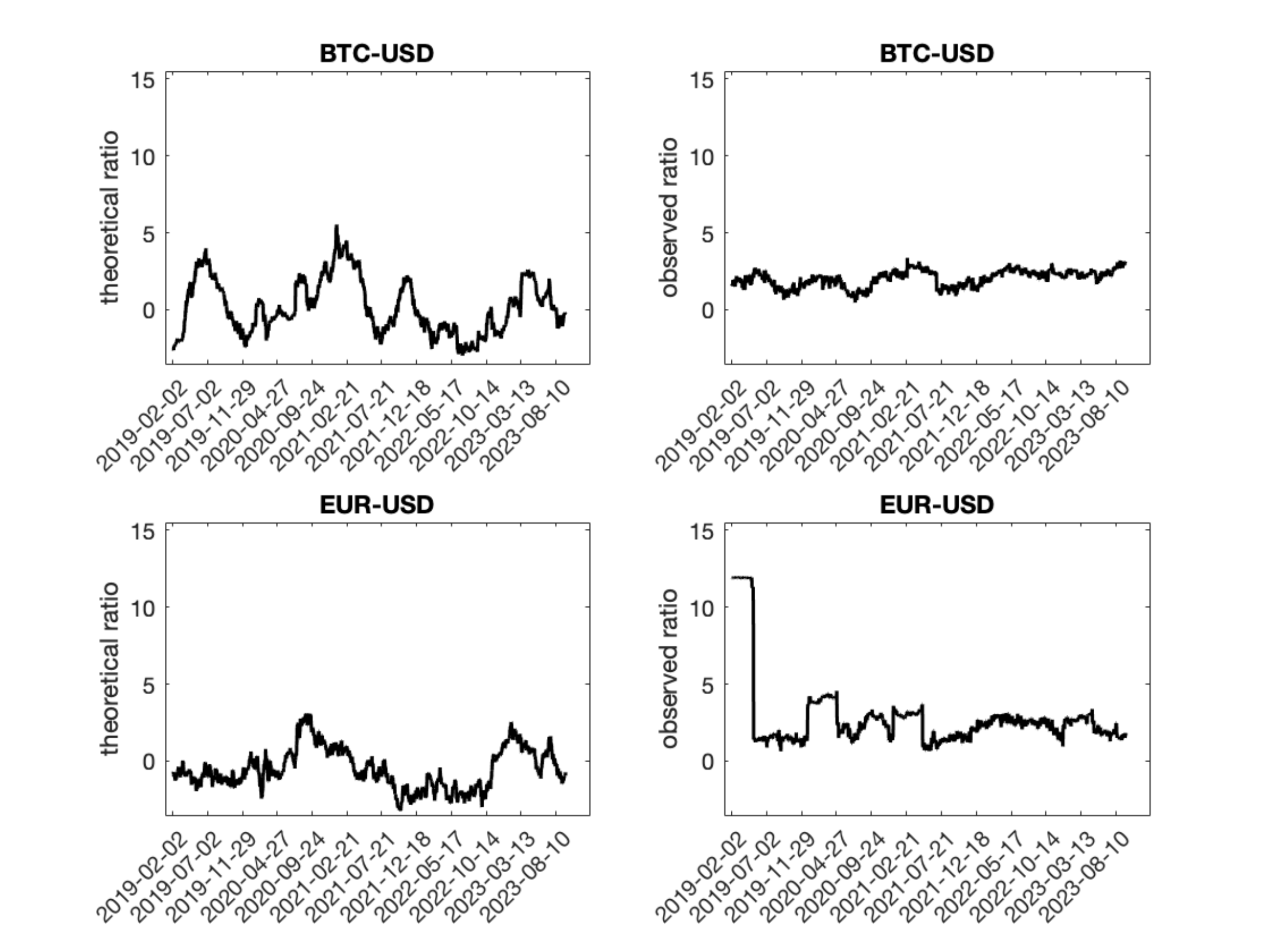}
\caption{\small  Comparison of ORPMM's theoretical drift-to-diffusion ratio of volume dynamics with observed values of the ratio. Linear regression with intercept only: BTC -- Intercept = 2.03, Standard deviation = 0.013; EUR -- Intercept = 2.81, Standard deviation = 0.057.
}\label{Ratio_BTC_EUR}
 \end{figure}

Upon examining Figure \ref{Ratio_BTC_EUR}, it is evident that the observed ratio does not coincide with the ORPMM's theoretical ratio. More interestingly, the observed ratio remains approximately constant throughout the considered time period.

We further assess the value of the observed ratio by conducting linear regression with intercept only, resulting in the following outcomes: BTC intercept equals 2.03 with a standard deviation of 0.013, and EUR intercept equals 2.81 with a standard deviation of 0.0567.

The drift-to-diffusion ratios are comparable with the ratios of PFE and VZ stocks rather than with those of the SPX and NA indices. 

In the left panels of Figure \ref{Risk_premium_BTC_EUR} we compare the volume risk premium $\psi+\rho\sigma^2$ (first column -- left panels) with the market risk premium, $\mu-r$ (second column -- left panels) while the right panels show the volume price of risk
$(\psi+\rho\sigma^2)/\sigma$ (first column -- right panels) and the market price of risk $(\mu-r)/\sigma$ (first column -- right panels).

Conducting a linear regression with intercept only for the volume risk yields the following results: the intercept for BTC-USD is 2.03 with a standard deviation of 0.013, while for EUR-USD, it is 2.981 with a standard deviation of 0.0567. Surprisingly, the reward for investing in BTC rather than EUR appears to be lower.

Regarding the market price of risk, we find an intercept of 0.7645 with a standard deviation of 0.0428 for the BTC-USD rate, and an intercept of -0.461 with a standard deviation of 0.032 for EUR-USD. The negative market price of risk for the EUR-USD exchange rate is consistent with the findings of the European Commission's Directorate General for Economic and Financial Affairs, as presented in their report on Euro-US Dollar Exchange Rate Dynamics (see specifically page 9: \url{https://economy-finance.ec.europa.eu/system/files/2020-11/eb055_en.pdf }).

\begin{figure}
\centering
\hspace{-1.cm}\includegraphics[scale=0.25]{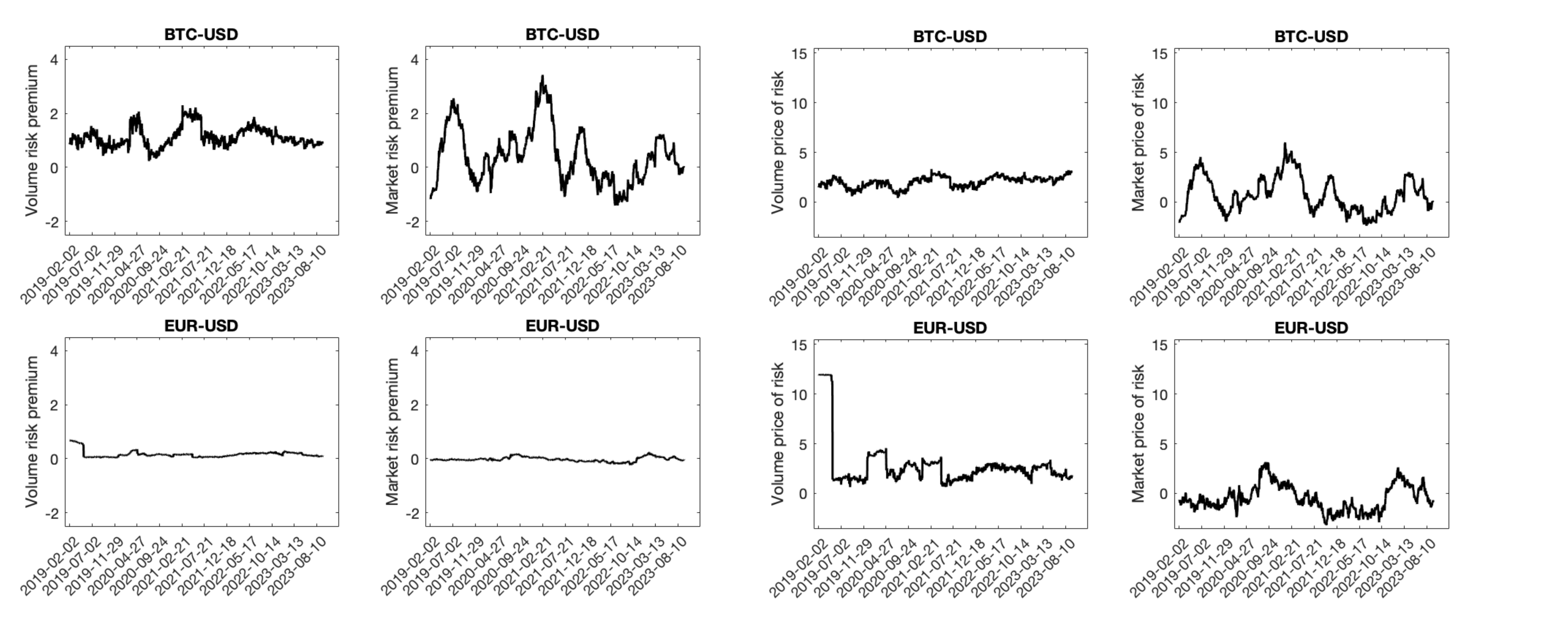}\caption{Volume risk premium, $\psi+\rho\sigma^2$ (first column, left panels), and the volume price of risk, $(\psi+\rho\sigma^2)/\sigma$ (first column, right panels).The market risk premium and the price of risk are in the second columns of the left and right panels, respectively. }\label{Risk_premium_BTC_EUR}
 \end{figure} 
 
Finally, we turn our attention to the trading price of risk in the ``ideal ORPMM world".
\begin{figure}
\centering
\includegraphics[scale=0.25]{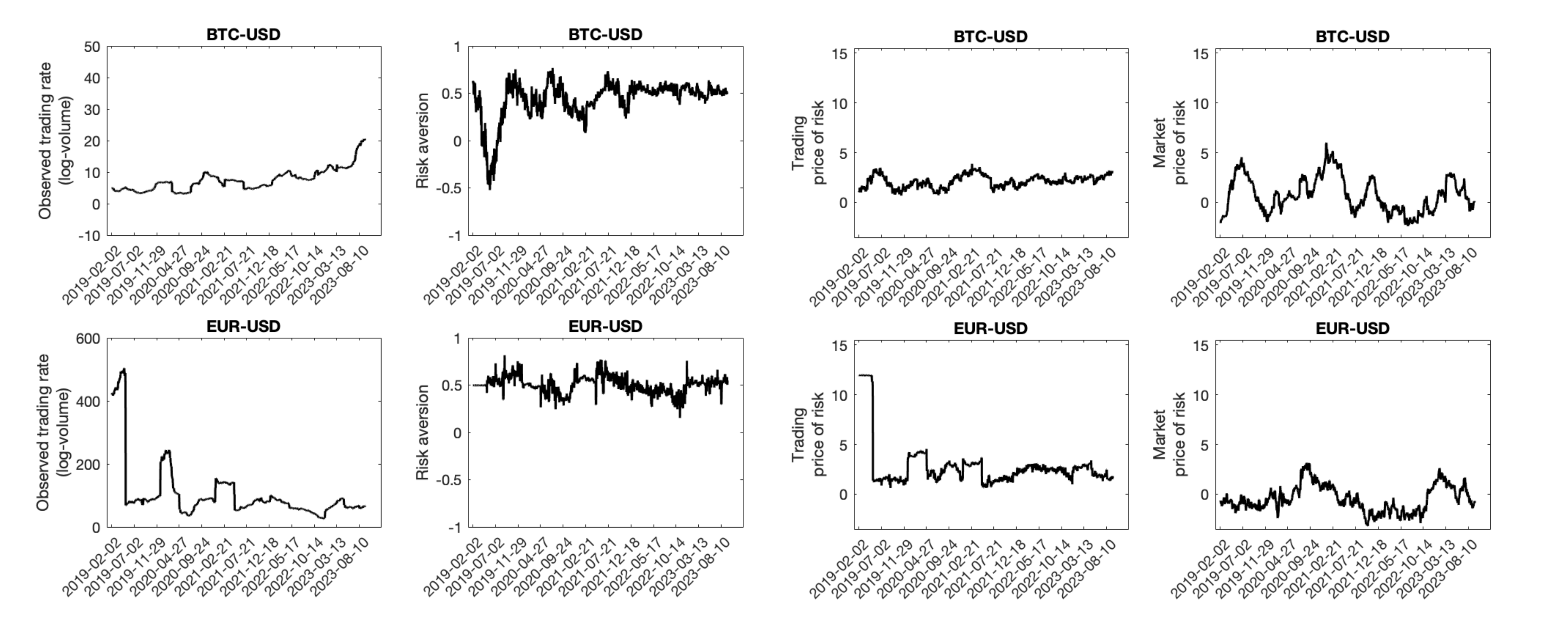}\caption{Trading intensity and risk aversion (left panels); trading price of risk and market price of risk (right panels).
} \label{Trading_risk_BTC_EUR}
 \end{figure} 
 In Figure \ref{Trading_risk_BTC_EUR}, the left panels display the estimated trading intensity $\eta$ and risk aversion $\bar{\gamma}$, while the right panels show the trading price of risk and market price of risk.
 
 We conducted linear regressions with intercept-only models for risk aversion, trading price of risk, and market price of risk, yielding the following results:\\
For the BTC-USD exchange rate: Risk aversion intercept = 0.436 (standard deviation = 0.00496), and trading price of risk intercept = 2.1303 (standard deviation = 0.01476).\\
For the EUR-USD exchange rate: Risk aversion intercept = 0.502 (standard deviation = 0.002184), and trading price of risk intercept = 2.805 (standard deviation = 0.0568).

We note that the trading price of risk closely resembles the volume price of risk, consistent with the findings of  \cite{KT1991}. It is noteworthy how risk aversion remains relatively constant over time for market makers operating in these assets.
The trading price of risk indicates that market makers trading BTC-USD and EUR-USD face a risk similar to Pfizer or Verizon stocks.

Finally, the negative market price of risk for the EUR-USD exchange rate is consistent with the findings of the European Commission's Directorate General for Economic and Financial Affairs, as presented in their report on Euro-US Dollar Exchange Rate Dynamics (see specifically page 9: \url{https://economy-finance.ec.europa.eu/system/files/2020-11/eb055_en.pdf }).

\subsection{Empirical analysis over a long horizon}

Let us conclude the empirical analysis by examining the trading intensity, risk aversion, and relevant risks (i.e., market price of risk, trading price of risk) over extended periods, including the very beginning of the asset.

We conducted analyses using daily price data for Pfizer from June 2nd, 1972, to May 26th, 2023, Verizon Communications from November 21st, 1983, to May 26th, 2023, the S\&P 500 from January 3rd, 1950, to April 8th, 2024, and BTC-USD from September 17th, 2014, to April 8th, 2024. We estimate the parameters using four years of daily data. Despite differing time spans, the results offer a unified interpretation, which we will elaborate on.

\begin{figure}[h!]
\centering
\hspace{-1.2cm}\includegraphics[scale=0.25]{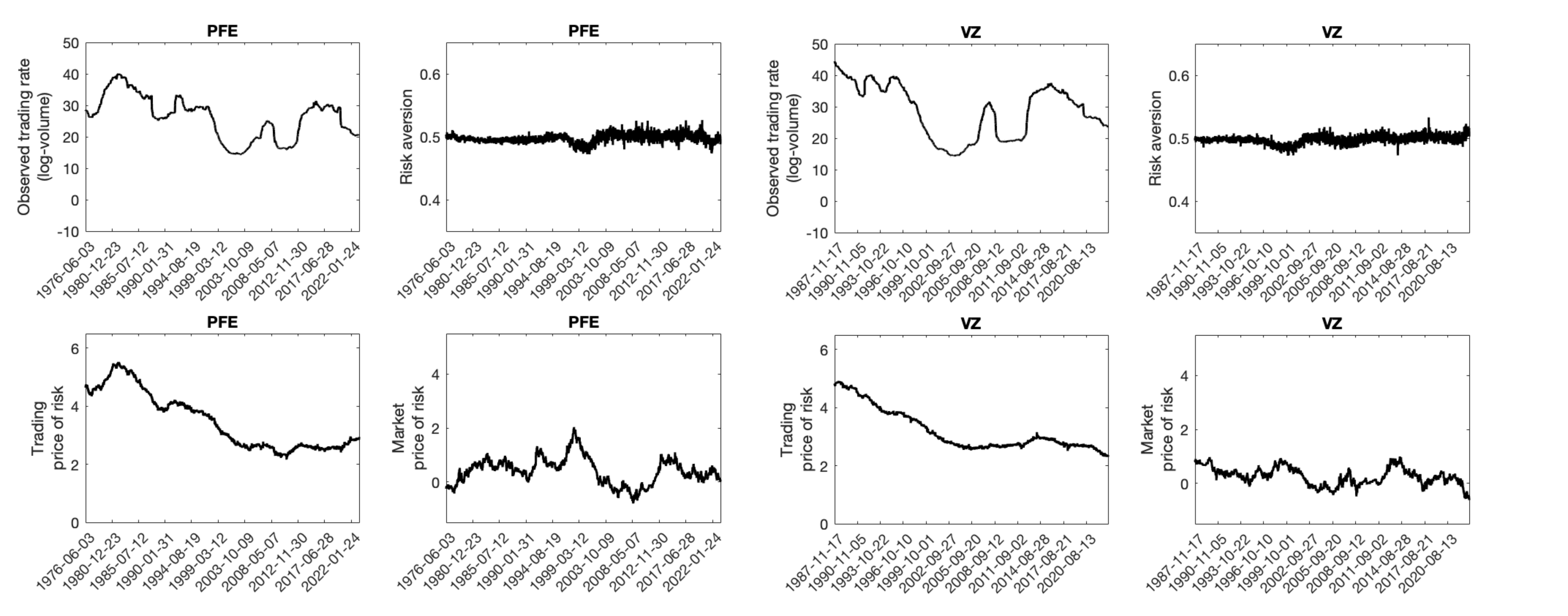}
\caption{Upper panels: trading intensity (left) and risk aversion (right). {Lower} panels: trading price of risk (left) and market price of risk (right) for PFE (left panels) and VZ (right panels)} \label{long_PFE_VZ}
 \end{figure} 
\begin{figure}[h!]
\centering
\hspace{-1.2cm}\includegraphics[scale=0.25]{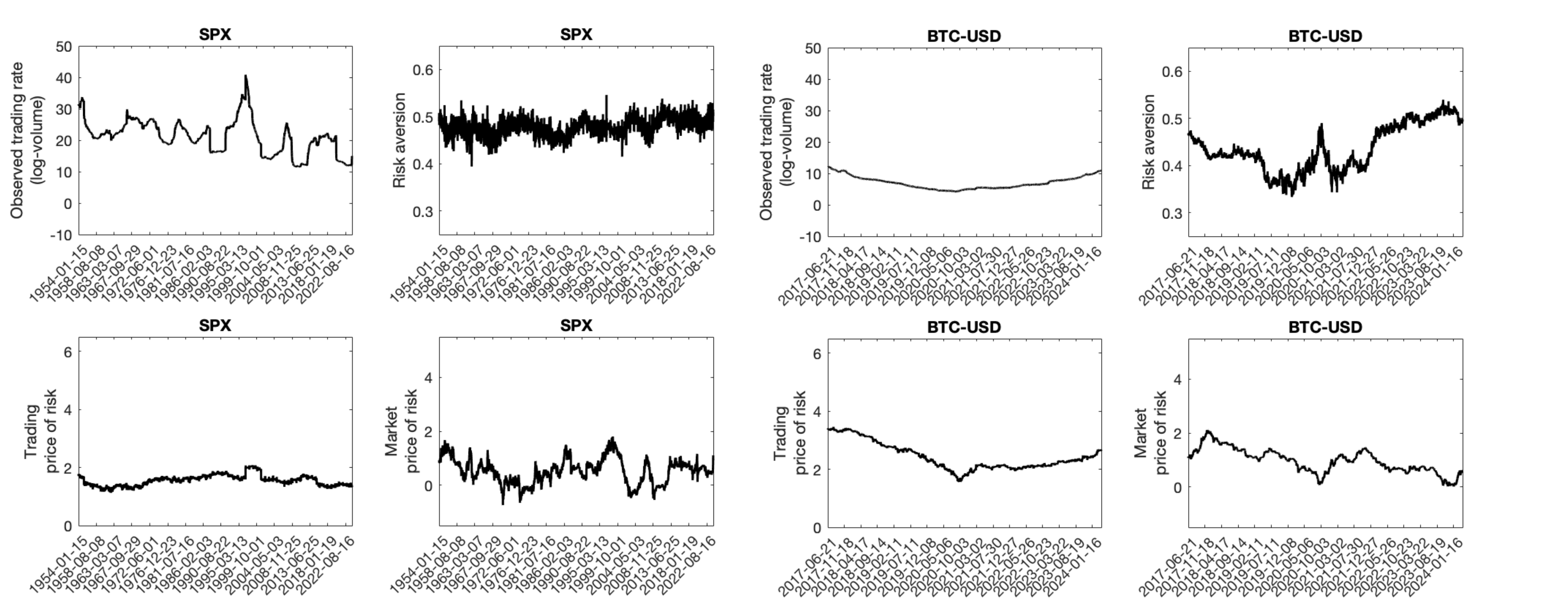}
\caption{Upper panels: trading intensity (left) and risk aversion (right). {Lower} panels: trading price of risk (left) and market price of risk (right).} \label{long_SPX_BTC}
 \end{figure} 

Figures \ref{long_PFE_VZ} and \ref{long_SPX_BTC} provide empirical evidence that, in the initial phases of an asset's life, the reward required for taking on investment risk is higher than 2-2.5, with the exception of the SPX index. However, risk aversion remains relatively constant at 0.5, except for the BTC-USD rate. In contrast, the market price of risk fluctuates across all the assets analyzed.

Interestingly, the trading price of risk serves as a measure of potential asset growth or market maker confidence in the asset. Higher values of the trading price of risk indicate lower market maker confidence but higher potential growth for the asset. Trading risk values exceeding 2.5 may suggest that the stock is not yet well-established.

In fact, the trading price of risk for Pfizer and Verizon Communications appears to be a decreasing function of time. On the other hand, the trading price of risk for the BTC-USD exchange rate decreases, reaching its minimum in November 2020, and then begins to rise again, reaching a value of approximately 2. This is likely due to the Bitcoin halving phenomenon. Bitcoin trading continues to exhibit a risk aversion of less than 0.5 up to the present day.

Finally, the S\&P 500 index shows a trading price of risk close to 2, indicating a very well-established asset.

\section{Conclusions}\label{conclusion}
{In this paper, we present a fully observable model for price-volume dynamics, and we show that the risky dollar volume, as generated by this model, can be interpreted as the {collective} behavior of market makers operating within a ORPMM frictionless framework. Additionally, we identify the impacted price dynamics as the one that ensures the observed total dollar volume managed by market makers is the solution to the ORPMM problem in a frictionless market.
In other words, market frictions are transferred into ``the price dynamics" to ensure that the observed dollar volume dynamics remain optimal in a frictionless market. The impacted price drift and volatility represent deviations from the asset's natural drift and volatility, caused by two sources of friction: the misalignment between the observed volume's drift-to-volatility ratio and the theoretical ORPMM ratio, as well as the misalignment between the price and volume Brownian motions.
Interestingly, the market impact model implied by the impacted price dynamics reveals two sources of risk: one arising from the risky asset and the other from the volume. Furthermore, the model enables the estimation of both market risk aversion and the price of risk implied by the impacted price (the trading price of risk). The findings of this paper suggest two future research directions: using the market impact model for price forecasting and scenario analysis, and applying the model in high-frequency settings with a stochastic trading intensity. Preliminary results on forecasting are presented in the online Appendix.}

\end{document}